\documentclass[preprint,3p]{elsarticle}
\usepackage{subfigure}
\usepackage[compat=1.1.0]{tikz-feynman}

\begin{document}
\begin{frontmatter}
\title{The role of multiplicative noise in critical dynamics}
\author[1,2]{Nathan O.  Silvano\corref{cor1}}
\ead{nathanosilvano@gmail.com }
\author[2]{Daniel G. Barci}
\ead{daniel.barci@gmail.com}
\address[1]{Center for Advanced Systems Understanding, Untermarkt 20, 02826 G\"orlitz, Helmholtz-Zentrum Dresden-Rossendorf, Bautzner Landstraße 400, 01328 Dresden}
\address[2]{Departamento de F{\'\i }sica Te\'orica, Universidade do Estado do Rio de Janeiro, Rua S\~ao Francisco Xavier 524, 20550-013, Rio de Janeiro, RJ,  Brazil} 
\cortext[cor1]{Corresponding author}

\begin{abstract}
We study the role of multiplicative stochastic processes in the description of the dynamics of an order parameter near a critical point.  We study equilibrium as well as  out-of-equilibrium properties.   By means of a functional formalism,  we build the Dynamical Renormalization Group equations for a real scalar order parameter with $Z_2$ symmetry,  driven by a class of multiplicative stochastic processes with the same symmetry. We  compute the flux diagram using a controlled  $\epsilon$-expansion,  up to order $\epsilon^2$. We find that, for dimensions $d=4-\epsilon$, the additive dynamic fixed point is unstable.  The flux runs to a {\em multiplicative fixed point} driven by a diffusion function $G(\phi)=1+g^*\phi^2({\bf x})/2$,  where $\phi$  is the order parameter and $g^*=\epsilon^2/18$ is  the fixed point value of the multiplicative noise coupling constant.  We show that, even though the position of the fixed point depends on  the stochastic prescription,  the critical exponents do not.   Therefore,  different dynamics driven by different stochastic prescriptions (such as It\^o, Stratonovich,  anti-It\^o and  so on)  are in the same universality class. 
 \end{abstract}

\end{frontmatter}

\tableofcontents

\newpage
%%%%%%%%%%%%%%%%%%%%%%%%%%%%%%%%%%
\section{Introduction}
Multiplicative stochastic processes have a long history and they were used to model very different dynamical systems~\cite{vanKampen,gardiner}.  Perhaps, the simplest example is the diffusion of a Brownian particle near a wall~\cite{Lancon2001,Lancon2002}.  Some other typical examples of multiplicative noise processes are  micromagnetic dynamics~\cite{GarciaPalacios1998, Aron2014, Arenas2018} and non-equilibrium transitions into absorbing states~\cite{Hinrichsen2000}.  
Moreover, multiplicative noise plays a central role in the description of out-of equilibrium systems~\cite{Grinstein1997, Munoz-2004}  and noise-induced phase transitions~\cite{Sancho1998,BarciMiguelZochil2016}.

In this paper, we study the role of multiplicative noise in the dynamics of phase transitions. 
Out-of-equilibrium evolution near continuous phase transitions is a fascinating subject. While equilibrium properties are strongly constrained by symmetry and dimensionality, the dynamics is much more involved  and it generally depends on conserved quantities and other details of the system. The interest in critical dynamics is rapidly growing up in part due to the  wide range of  multidisciplinary applications in which criticality  has deeply impacted.  For instance, the collective behavior of  different biological systems has critical properties,  displaying  space-time correlation functions with nontrivial scaling laws~\cite{TGrigera-2019,Gambassi-2021}. Other interesting examples come from epidemic spreading models where  dynamic percolation  is observed near multicritical  points~\cite{Janssen-2004}. Moreover,   strongly correlated systems, such as antiferromagnets in transition-metal oxides~\cite{Cabra-2005,Bergman-2006},  usually described by  dimmer models or related quantum field theory models~\cite{Hsu-2013}, seem to have  anomalous critical dynamics~\cite{Isakov-2011}. 

The standard approach to critical dynamics is the ``Dynamical Renormalization Group (DRG)''~\cite{Bausch-1976},  distinctly developed  in  a seminal paper by Hohenberg and Halperin~\cite{Hohenberg-Halperin-1977}.  The simplest starting point is to assume that,  very near a critical point,  the dynamics of the order parameter is governed by a dissipative process driven by  an overdamped {\em additive noise Langevin equation}.  The typical relaxation time near a fixed point is  given by  $\tau\sim \xi^z$,  where $\xi$ is the correlation length and $z$ is the dynamical critical exponent.  At a critical point, $\xi\to\infty$ and therefore $\tau\to \infty$, meaning that the system does not reach the equilibrium at criticality.  Together with usual static exponents,  $z$ defines the universality class of the transition.  Interestingly,  since the symmetry of the model does not constrain dynamics,  there are different dynamic universality classes for the same critical point.

As usual in Renormalization Group (RG)  theory~\cite{Cardy-book}, 
a RG transformation generates all kind of interactions,  compatible with the symmetry of the system.  For this reason, a consistent study of a RG flux should begin, at least formally, with the most general Hamiltonian containing all couplings compatible with symmetry.  Interestingly,  DRG transformations not only generate new couplings in the Hamiltonian,  but also modify the probability distribution of the stochastic process initially assumed.  Thus,  the stochastic noise probability distribution also flows with the DRG transformation, {\em i.e.},  it is scale dependent. 
In particular, we will show that one-loop perturbative corrections generate couplings compatible with multiplicative noise stochastic processes,   even in the case of assuming an additive dynamics as a starting condition.  In order to understand the fate of these couplings,  we decided to  analyze a more general  dynamics for the order parameter near criticality. We  assume a dissipative process driven by a {\em  multiplicative noise Langevin equation}.  
For concreteness,  we address a simple model of a non conserved real scalar order parameter, $\phi({\bf x},t)$ with quartic coupling $\phi^4({\bf x},t)$ (model A of Ref.~\cite{Hohenberg-Halperin-1977}).  The dynamics is driven by  a  multiplicative noise Langevin equation,  characterized by a  general  dissipation function $G(\phi)$,  with the same symmetry of the Hamiltonian.   

Initially,  we analyze the influence of  a stochastic multiplicative evolution on the equilibrium properties of the system.  For this,  we compute the asymptotic stationary probability distribution.   We show that  the multiplicative diffusion function modifies the asymptotic equilibrium potential.     It is  worth to  note that,  depending on the parameters of the model,  multiplicative dynamics could have dramatic effects on thermodynamics,  possibly changing  the order  of the phase transition.  

To study the dynamics of this problem, we implement a DRG  using a functional formalism.   For spatial dimensions greater than the upper critical one ($d_c=4$ in the example discussed),     the Gaussian fixed point controls the physics of the phase transition and all multiplicative noise couplings are irrelevant.  Thus, the assumption of an overdamped Langevin additive dynamics is correct.  In this case,  the static critical exponents are provided as usual by a Landau theory and the dynamical critical exponent $z=2$.    However,   for $d<4$,  fluctuations are important,  the Gaussian fixed point is unstable and the problem of the fate of  multiplicative couplings is more involved.    
We perform a systematic well controlled $\epsilon=4-d$ expansion of the DRG up to  order $\epsilon^2$. 
We find that,  at order $\epsilon$,   the Wilson-Fisher fixed point~\cite{Wilson-1972}  shows up, the multiplicative couplings are irrelevant and the dynamical critical exponent $z=2+O(\epsilon^2)$, in agreement with the results of Ref.~\cite{Hohenberg-Halperin-1977}.   However,  at order $\epsilon^2$,   the additive dynamics is no longer a fixed point of the DRG equations,   giving  place to a novel fixed point with a true multiplicative dynamics,  codified by the diffusion function $G(\phi)=1+g^* \phi^2/2$,  where $g^*\sim \epsilon^2$ is a fixed value of the coupling constant that characterizes the multiplicative dynamics.

The paper is organized as follows: In \S \ref{Sec:Multiplicative-Noise} we  make a brief review of multiplicative stochastic processes, beginning with a single variable,  passing through multiple variable systems,  ending with extended continuous systems appropriated to describe order parameters and phase transitions.   In this section,  we analyze in detail equilibrium properties.  In \S \ref{Sec:Formalism} we describe the functional formalism to describe dynamics of an order parameter,  presenting an specific model.  In section \ref{Sec:DRG},  we built up the DRG equations and compute the flux diagram.  Finally,  we discuss our results in section \ref{Sec:Discussions}.
We leave calculation details  for the appendices \ref{App:DRG-equations}, \ref{App:g2},  \ref{App:Integrals} and \ref{App:Rescaling}.

%%%%%%%%%%%%%%%%%%%%%%%%%%%%%%%%%%%%
\section{Brief review of multiplicative stochastic processes}
\label{Sec:Multiplicative-Noise}
%%%%%%%%%%%%%%%%%%%%%%%%%%%%%%%%%%%%
To make the paper self-contained and to establish notation,  we present  in this section a very brief review of multiplicative stochastic processes.  Although the content of this section is not completely novel, we believe that the way to present the subject clarifies some usual misunderstandings in the literature. 
%%%%%%%%%%%%%%%%%%%%
\subsection{Single stochastic variable}
\label{subsec:SingleVariable}
Let us consider a single stochastic variable $\phi(t)$,  whose dynamics is driven by the multiplicative Langevin equation
\begin{equation}
\frac{d\phi(t)}{d t}=F[\phi]+ G(\phi)\eta(t)\; , 
\label{eq:Langevin-SingleVariable}
\end{equation}
where $\eta(t)$ is a Gaussian white noise
\begin{eqnarray}
\langle \eta(t) \rangle&=&0 \; , 
\label{eq:eta} \\
\langle \eta(t)\eta(t') \rangle &=&\beta^{-1}\delta(t-t')\; .
\label{eq:etaeta}
\end{eqnarray}
$F(\phi)$ is an arbitrary drift force and $G(\phi)$ is a diffusion  function that characterizes the multiplicative stochastic dynamics.   The constant $\beta$ measures the noise intensity and,  under specific equilibrium conditions, can be interpreted as a temperature parameter $\beta=1/k_B T$ ($k_B$ is the Boltzmann constant).

In order to completely define Eq.~(\ref{eq:Langevin-SingleVariable}),  it is necessary to fix the stochastic prescription necessary to define the Wiener integrals.  Along this paper we use the {\em Generalized Stratonovich prescription},  also known as $\alpha$-prescription, which is labeled by a real parameter $0\le \alpha\le 1$.   For concreteness,   a solution of Eq. (\ref{eq:Langevin-SingleVariable}) will contain integrals of the type
\begin{equation}
\int G(\phi(t)) \eta(t) dt=  \int G(\phi(t)) dW
\end{equation}
where $W(t)$ is a Wiener process.   By definition, the Riemann-Stieltjes integral is 
\begin{equation}
 \int G(\phi(t)) dW=\lim_{n\to \infty}\sum_{i=1}^n G(\phi(\tau_i))\left[W(t_{i+1})-W(t_i)\right]
\end{equation}
where $\tau_i$ is taken in the interval $[t_i,t_{i+1}]$.  The limit $n\to \infty$ is taken in the mean quadratic sense~\cite{vanKampen,gardiner}.
For smooth measures $W(t)$,  the integral does not depend on  $\tau_i$.  However,   since it is a Wiener process,  the limit depends on the prescription to choose $\tau_i$.  In the  {\em Generalized Stratonovich prescription},  $\tau_i$ is fixed  in the following way: 
\begin{equation}
G(\phi(\tau_i))=G[(1-\alpha) \phi(t_i)+\alpha \phi(t_{i+1})]
\end{equation}
where $0\le \alpha\le 1$.   In this way,  $\alpha=0$ corresponds with  the pre-point  It\^o prescription~\cite{Ito}, $\alpha=1/2$ is the Stratonovich one~\cite{Stratonovich}, 
while $\alpha=1$ is known as the  H\"anggi-Klimontovich or anti-It\^o (post-point) convention~\cite{Hanggi1982,Klimontovich}.  
Therefore, the dynamics described by equation~\ref{eq:Langevin-SingleVariable} is completely determined by the drift force $ F(\phi)$, the diffusion function $G(\phi)$ and the stochastic prescription $\alpha$. 

It is possible to have a deeper insight by looking at  the Fokker-Planck equation for the probability distribution $P(\phi,t)$.  It can be cast in the form of a continuity equation~\cite{Arenas2012-2} 
\begin{equation}
\frac{\partial P(\phi,t)}{\partial t}+\frac{\partial J(\phi,t)}{\partial \phi}=0
\label{eq:FP}
\end{equation}
where the probability current is given by 
\begin{equation}
J(\phi,t)=\left[ F(\phi)-(1-\alpha)\beta^{-1} G(\phi)G'(\phi)\right] P(\phi,t)-\frac{1}{2}\beta^{-1}G^2(\phi)\frac{\partial P(\phi,t)}{\partial \phi}
\label{eq:J}
\end{equation}
in which $G'=dG/d\phi$.
It is worth noting  that for each value of $\alpha$, we have  different dynamics.  This effect is due to multiplicative noise.  In fact, note that the term proportional to $\alpha$  is always multiplied by $dG/d\phi$.  Therefore,  for additive noise,   $dG/d\phi=0$  and,  as a consequence,  the dynamics does not depend on the stochastic prescription. 

From equation (\ref{eq:FP}),  it is immediate to access the equilibrium properties of the stochastic system.   Assuming that, at long times,  the system converges to a stationary state,  the stationary probability
\begin{equation}
P_{\rm st}(\phi)=\lim_{t\to\infty} P(\phi,t)
\end{equation}
satisfies, 
\begin{equation}
\frac{d J_{\rm st}}{d\phi}=0  \; ,
\end{equation}
where $J_{\rm st}(\phi)=\lim_{t\to \infty} J(\phi,t)$ is given by equation (\ref{eq:J}), evaluated at the stationary probability $P_{\rm st}(\phi)$.
The equilibrium solution of the Fokker-Planck  equation satisfies $J_{\rm st}(\phi)=0$.  We immediately obtain 
\begin{equation}
P_{eq}(\phi)={\cal N} e^{-\beta U_{\rm eq}(\phi)}
\label{eq:Peq}
\end{equation}
where ${\cal N}$ is a normalization constant and 
\begin{equation}
U_{\rm eq}(\phi)=-2 \int^{\phi} \frac{F(x)}{G^2(x)}dx+(1-\alpha)\beta^{-1}  \ln\left(G^2(\phi)\right)
\label{eq:UeqF}
\end{equation}
Eqs. (\ref{eq:Peq}) and (\ref{eq:UeqF}) codify the asymptotic equilibrium properties of the stochastic system driven by the Langevin equation (\ref{eq:Langevin-SingleVariable}).    The equilibrium properties are determined by both functions $\{F,G\}$ and the stochastic prescription $\alpha$.  If the stochastic system represents a conservative system in a noisy environment,  {\em i.e.},  if it is possible to identify a deterministic system defined by a potential or a Hamiltonian $H(\phi)$,  then $F(\phi)$ is time independent and can be written in terms of the Hamiltonian as  
\begin{equation}
F(\phi)=-\frac{1}{2}G^2(\phi)\frac{d H}{d\phi} \; .
\label{eq:ER}
\end{equation}
This equation is a generalization of the  Einstein relation~\cite{Arenas2012-2}.
In this case, the equilibrium potential, Eq. (\ref{eq:UeqF}) takes the simpler form, 
\begin{equation}
U_{\rm eq}(\phi)=H(\phi)+\frac{(1-\alpha)}{\beta}  \ln\left(G^2(\phi)\right)
\label{eq:Ueq}
\end{equation}
As expected,  the equilibrium potential not only depends  on  the Hamiltonian  and the diffusion function, but also on the stochastic prescription $\alpha$.     Notice that,  for $\alpha=1$,   $U_{\rm eq}=H$ and the probability distribution is  the Boltzmann distribution, as it should be for a thermodynamic physical system.  For this reason,   the $\alpha=1$ prescription is also known as {\em thermal prescription}.    For any other value of $\alpha$, even the more usual  Stratonovich one, $\alpha=1/2$, the equilibrium potential is modified by the diffusion function $G$.  This correction is proportional to $\beta^{-1}$,   since it has a noisy origin.  

An important consequence of Eq. (\ref{eq:Ueq}) is that in a zero dimensional stochastic system (just one variable $\phi(t)$), the equilibrium state is always reached for almost any reasonable $G(\phi)$. The only condition is that the equilibrium distribution should be normalizable, {\em i.e.}, ${\cal N}^{-1}=\int d\phi \exp\{-\beta U_{\rm eq}\}$ should be finite.

%%%%%%%%%%%%%%%%%%%%%%%%%%%%%%%%%%%%
\subsection{Multiple stochastic variables}
\label{subsec:MultipleVariables}
The generalization of the single variable stochastic dynamics to a multiple variable system is straightforward. However, 
the equilibrium properties are quite different in this last case~\cite{Miguel2015}. Let us consider a generalization of Eq. (\ref{eq:Langevin-SingleVariable}) to multiple variables written in the following way:
\begin{equation}
\frac{d\phi_i(t)}{d t}=F_i[\phi_1,\ldots,\phi_N]+ G_{ij}(\phi_1,\ldots,\phi_N)\eta_j(t)\; , 
\label{eq:Langevin-FG-V}
\end{equation}
where $\phi_i$, with $i=1,\dots,n$,   are $n$ independent stochastic variables and   $\eta_j(t)$ with $j=1,\dots,m$ are $m$ Gaussian white noise processes,
\begin{eqnarray}
\langle \eta_i(t) \rangle&=&0 \; , 
\label{eq:etaV} \\
\langle \eta_i(t)\eta_j(t') \rangle &=&\beta^{-1}\delta_{ij}\delta(t-t')\; .
\label{eq:etaetaV}
\end{eqnarray}
$F_i(\phi)$ is an $n$ component drift force and $G_{ij}(\phi)$ is a diffusion matrix.  The constant $\beta$ measures the noise intensity and,  under very specific conditions that we will carefully discuss, can be interpreted as a temperature parameter $\beta=1/k_B T$.

The Fokker-Planck equation can be cast in the form of a continuity equation
\begin{equation}
\frac{\partial P(\phi,t)}{\partial t}+\partial_i  J_i(\phi,t)=0 \; .
\label{eq:FPV}
\end{equation}
We are using the notation $\partial_i\equiv \partial~/\partial \phi_i$ and we are assuming summation over repeated indexes.
The probability current  is given by 
\begin{equation}
J_i(\phi,t)=\left[ F_i(\phi)+\alpha\beta^{-1} G_{k\ell}(\phi)\partial_k G_{i\ell}(\phi)\right] P(\phi,t)-\frac{1}{2}\beta^{-1}\partial_j\left( G_{i\ell}(\phi)G_{j\ell}(\phi) P(\phi,t)\right)
\label{eq:JV}
\end{equation}
 This equation is formally equivalent to Eq.(\ref{eq:J}). The main differences are that, in the one hand, it  is an $n$ component vector equation and, on the other hand, it has a more involved matrix structure of the diffusion function. 

The stationary solutions should satisfy 
\begin{equation}
\partial_i J^{\rm st}_i(\phi)=0
\end{equation}
Depending on $F_i$ and $G_{ij}$,  there could be several solutions of this equation, representing non-equilibrium steady states.   
The  true equilibrium solution, if it exists,  is given by  
\begin{equation}
J^{\rm st}_i(\phi)=0\;. 
\label{eq:Jieq}
\end{equation}
A natural question is: which are the conditions for the existence of equilibrium? 
Assuming, as in the one dimensional case, a stationary distribution of the type 
\begin{equation}
P_{\rm eq}(\phi_1,\ldots,\phi_n)={\cal N} e^{-\beta U_{\rm eq}(\phi_i,\ldots,\phi_n)}
\label{eq:PeqMV}
\end{equation}
and replacing Eq. (\ref{eq:PeqMV}) into Eq. (\ref{eq:Jieq}), 
we find for the equilibrium potential, 
\begin{equation}
\partial_p U_{\rm eq}=-2 G^{-1}_{sp}G^{-1}_{si} F_i+\beta^{-1}\left\{G^{-1}_{sp}\partial_jG_{js}-
\alpha G^{-1}_{sp}G^{-1}_{si}G_{k\ell}\partial_kG_{i\ell}     \right\}\; .
\label{eq:partialU}
\end{equation}

For constant $G_{ij}$ (additive noise), we expect that the equilibrium potential should coincide with the Hamiltonian of the system. For this, we need to impose a generalized Einstein relation 
\begin{equation}
F_i(\phi)=-\frac{1}{2}G_{i\ell}G_{\ell k} \partial_k H \; .
\label{eq:ERi}
\end{equation}
In this way, Eq. (\ref{eq:partialU}) takes the simpler form, 
\begin{equation}
\partial_p U_{\rm eq}=\partial_p H +\beta^{-1 }A_p(G) \; , 
\label{eq:Ueqi}
\end{equation}
where the vector $A_p(G)$ is given by
\begin{equation}
A_p(G)=G^{-1}_{sp}\partial_j G_{js}+(1-2\alpha) G^{-1}_{sp}G^{-1}_{si} G_{k\ell}\partial_k G_{i\ell}
\label{eq:A}
\end{equation}
Notice that $A_p$ depends on derivatives of the diffusion functions. Thus, for additive noise, $A_p=0$ and from Eq. (\ref{eq:Ueqi}), 
$U_{\rm eq}=H$ as expected. 

However, for multiplicative noise, there are no solutions of Eq. (\ref{eq:Ueqi})  for  a general diffusion matrix $G_{ij}(\phi)$.
A sufficient condition to solve Eq. (\ref{eq:Ueqi}) is that $A_p$ should be written as a gradient in the form
\begin{equation}
A_p(\phi)=\partial_p\Delta(\phi) \;,
\label{eq:Gradient}
\end{equation}
where $\Delta(\phi)$ is a scalar function of $\phi_i$, with $i=1,\ldots,n$.
In this case, the solution reads, 
\begin{equation}
U_{\rm eq}(\phi)= H(\phi)+\beta^{-1}\Delta(\phi)
\label{eq:UeqMultivariable}
\end{equation}

The condition (\ref{eq:Gradient}) is satisfied if  the following relation holds: 
\begin{equation}
\partial_{i_1} \tilde A_{i_1,\ldots,i_{n-1}}=0 
\label{eq:divergentA}
\end{equation}
where we have introduced the dual tensor 
\begin{equation}
\tilde A_{i_1,\ldots,i_{n-1}}=\epsilon_{i_1,\ldots,i_{n}} A_{i_n}
\label{eq:dual}
\end{equation}
and $\epsilon_{i_1,\ldots,i_{n}}$ is the completely antisymmetric Levi-Civita tensor in $n$-dimensions.

Of course, for a single variable ($n=1$), $\Delta=(1-\alpha)\ln G^2$, and  Eq. (\ref{eq:UeqMultivariable}) reduces to Eq. (\ref{eq:Ueq}) as it should be. 

Summarizing, not any multiplicative stochastic process drives the system to an equilibrium state. A  sufficient condition for the existence of equilibrium is that $G_{ij}(\phi)$ should satisfy  Eq. (\ref{eq:divergentA}). With this condition, the equilibrium potential is given by Eq. (\ref{eq:UeqMultivariable}).  Some particular cases of this expression  were explored in Ref.~\cite{Miguel2015}.
Moreover, a class  of non-equilibrium stationary solutions and related non-equilibrium phase transitions where presented in Ref. ~\cite{BarciMiguelZochil2016}.

%%%%%%%%%%%%
\subsection{Continuum systems}
\label{subsec:Extended}
Phase transitions are characterized by the behavior of an order parameter and its fluctuations. Thus, to describe dynamics of phase transitions it is necessary to deal with continuum systems with infinite degrees of freedom. This type of system is a special case of the multiple-variable systems described in the last subsection.  
A  multiplicative  Langevin equation for an order parameter  can be written as
\begin{equation}
\frac{\partial\phi({\bf x}, t)}{\partial t}=F\left(\phi({\bf x},t)\right)+ G(\phi({\bf x}))\eta({\bf x}, t)\; , 
\label{eq:Langevin-Extended}
\end{equation}
where $\phi({\bf x},t)$ is a scalar field in $d$-dimensions and  $\eta({\bf x}, t)$ is a Gaussian white noise field
\begin{eqnarray}
\langle \eta({\bf x},t) \rangle&=&0 \; , 
\label{eq:etaxt} \\
\langle \eta({\bf x}, t)\eta({\bf x'}, t') \rangle &=&\beta^{-1}\delta(t-t')\delta^d({\bf x}-{\bf x'})\; .
\label{eq:etaetaxt}
\end{eqnarray}
$F(\phi({\bf x},t))$ is an arbitrary drift force and  $G(\phi({\bf x}))$ is a diffusion  function that characterizes the multiplicative stochastic dynamics.  

The Fokker-Planck equation is written as, 
\begin{equation}
\frac{\partial P(\phi({\bf x}), t)}{\partial t}+\int d^d x\frac{\delta J(\phi, t)}{\delta\phi({\bf x})}=0\; , 
\label{eq:FPex}
\end{equation}
where $\delta/\delta\phi({\bf x})$ is a functional derivative and the probability current is now given by 
\begin{equation}
J(\phi({\bf x}),t)=\left[ F(\phi({\bf x}))-(1-\alpha)\beta^{-1} G(\phi({\bf x}))\frac{\delta G(\phi)}{\delta \phi({\bf x})}\right] P(\phi({\bf x}),t)-\frac{1}{2}\beta^{-1}G^2(\phi({\bf x}))\frac{\delta P(\phi,t)}{\delta \phi({\bf x})}\; .
\label{eq:Jex}
\end{equation}
The equilibrium solution of this equation takes the exponential form, 
\begin{equation}
P_{eq}(\phi)={\cal N} e^{-\beta U_{\rm eq}[\phi({\bf x})]}\; .
\label{eq:Peqex}
\end{equation}
If we assume that the deterministic system is given by a Hamiltonian $H[\phi({\bf x})]$,  then the Einstein relation reads 
\begin{equation}
F(\phi)=-\frac{1}{2}G^2(\phi({\bf x}))\frac{\delta H}{\delta\phi({\bf x})} \; .
\label{eq:ERex}
\end{equation}
Solving the equation $\lim_{t\to\infty} J(\phi({\bf x}),t)=0$, we find for the equilibrium potential
\begin{equation}
U_{\rm eq}[\phi]=H[\phi({\bf x})]+\frac{(1-\alpha)}{\beta} \int d^dx \ln\left(G^2(\phi({\bf x}))\right)\; .
\label{eq:Ueqex}
\end{equation}
The structure of  $U_{\rm eq}$ is  similar  to the single variable case (Eq. (\ref{eq:Ueq})), due to the diagonal form of the diffusion matrix; the Hamiltonian acquires a logarithmic correction proportional to $(1-\alpha)$.

%%%%%%%%%%%%%%%%%%%%%%%%%%%%%%%%%%%%
\subsection{Equilibrium properties}
\label{subsec:Equilibrium}
%%%%%%%%%%%%%%%%%%%%%%%%%%%%%%%%%%%%
Equilibrium properties  are ruled by the potential $U_{\rm eq}[\phi]$ given by equation (\ref{eq:Ueqex}).  In general, the equilibrium distribution is not of the Boltzmann type. In fact, by replacing Eq. (\ref{eq:Ueqex}) into Eq.  (\ref{eq:Peqex}), it is immediate to realize
\begin{equation}
P_{\rm eq}[\phi]={\cal N} \det\left(G^{-2(1-\alpha)}[\phi]\right) e^{-\beta H[\phi]} \; .
\label{eq:Peqdistr}
\end{equation}
Thus, there is a nontrivial prefactor coming from the diffusion function $G(\phi)$ multiplying the usual Boltzmann distribution. 
It is important to note that, similarly to the zero dimensional case, the prefactor depends on the stochastic prescription. In particular, in the anti-It\^o prescription $\alpha=1$, the multiplicative noise contributions to the equilibrium distribution disappears, and the probability distribution is of the usual Boltzmann type. 

In order to explore the effect of the multiplicative noise on the equilibrium properties of the system, let us consider a simple example.  Consider a mean field Hamiltonian invariant under the transformation $\phi\to -\phi$, 
\begin{equation}
H=\frac{r}{2}\phi^2+\frac{u}{4!}\phi^4+v \phi^6
\end{equation}
and a diffusion function 
with the same symmetry of the Hamiltonian, 
\begin{equation}
G=\sqrt{1+g\phi^2}\; ,
\end{equation}
where  $\phi=\langle\phi(x)\rangle$ is a uniform order parameter and $g>0$ is the multiplicative noise coupling constant.  The  equilibrium potential per unit volume reads,  
\begin{equation}
U_{\rm eq}=\frac{r}{2}\phi^2+\frac{u}{4!}\phi^4+ v\phi^6 +\frac{(1-\alpha)}{\beta} \ln\left(1+g \phi^2\right)
\end{equation} 
The case  $u,v>0$ and $g=0$ is the usual Landau expansion of the free energy of the uniaxial ferromagnet. Indeed, for $r>0$, 
the system is in a disordered phase ($\phi=0$) and for $r<0$, the system is in a  symmetry broken ordered phase ($\phi\neq 0$). Thus, 
the critical temperature is at $r=0$. Very near the transition, $\phi\sim 0$.  Thus, in order to explore the effects of multiplicative noise ($g\neq 0$), we can expand the logarithm in powers of $g\phi^2$, obtaining an approximated equilibrium potential, 
\begin{equation}
U_{\rm eq}\sim \frac{\tilde r}{2}\phi^2+\frac{\tilde u}{4!}\phi^4+ \tilde v \phi^6+\ldots
\label{eq:UeqCorrected}
\end{equation} 
where 
\begin{eqnarray}
\tilde r&=&r+ 2  \frac{(1-\alpha)}{\beta} g  
\label{eq:rtilde} \\
\tilde u&=&  u-12    \frac{(1-\alpha)}{\beta} g^2 
\label{eq:utilde}\\
\tilde v&=& v+  \frac{1}{3}\frac{ (1-\alpha)}{\beta}  g^3
\label{eq:vtilde}
\end{eqnarray}
 The corrections to $r$ and $v$ are always positive, while the correction to $u$ is negative. 
Thus, if 
$
g^2/\beta<u/(12 (1-\alpha))
$ 
 then $\tilde u>0$. In this case, the phase transition is still a second order one, albeit with a reduced critical temperature. 
 On the other hand, if
 $
g^2/\beta>u/(12 (1-\alpha))
$ 
then, $\tilde u<0$ and the transition is no longer a second order one, but a first order phase transition. 
Therefore, depending on the intensity of the noise $\beta$, and the multiplicative coupling constant $g$, the effects of multiplicative noise on the equilibrium properties could be quite dramatic. Indeed, it could change the order of the phase transition. 
For small values of the noise or the multiplicative coupling constant, the transition remains second order and we expect that fluctuations do not change the universality class. However, the out-of-equilibrium properties of the system could be substantially affected by multiplicative noise. This is the main point that we will address in the following.

%%%%%%%%%%%%%%%%%%%%%%%%%%%%%%%%%%%%
\section{Functional formalism}
\label{Sec:Formalism}
%%%%%%%%%%%%%%%%%%%%%%%%%%%%%%%%%%%%
The dynamical evolution that converges, at very long times, to the equilibrium potential of Eq. (\ref{eq:Ueqex}) is 
driven by the Langevin equation, 
\begin{equation}
\frac{\partial\phi({\bf x},t)}{\partial t}=-\frac{\Gamma}{2} G^2(\phi)\frac{\delta H}{\delta \phi({\bf x},t)}+ G(\phi)\eta({\bf x},t)\; , 
\label{eq:Langevin}
\end{equation}
where $\eta({\bf x}, t)$ is a Gaussian white noise field,
\begin{eqnarray}
\langle \eta({\bf x},t) \rangle&=&0 \; , 
\label{eq:etax} \\
\langle \eta({\bf x},t)\eta({\bf x}',t') \rangle &=&\frac{\Gamma}{\beta}\delta({\bf x}-{\bf x}')\delta(t-t')\; .
\label{eq:etaxetax}
\end{eqnarray}
$\Gamma$ is a diffusion constant  and $\langle\ldots\rangle$ represents stochastic mean value.

The observables of the model are codified in the dynamical $n$-point correlation functions $\langle\phi({\bf x}_1, t_1)\ldots\phi({\bf x}_n, t_n)\rangle_\eta$ of the order parameter,  where the stochastic expectation values should be computed by using the Langevin Eq.  (\ref{eq:Langevin}),  with the white noise distribution,  Eqs. (\ref{eq:etax}) and (\ref{eq:etaxetax}).  From a theoretical perspective,  it is more convenient to use a functional formalism in which we define a generating functional $Z(J)$, in such a way that  the correlation functions
\begin{equation}
\langle\phi({\bf x}_1, t_1)\ldots\phi({\bf x}_n, t_n)\rangle_\eta= \left.\frac{\delta^n Z(J)}{\delta J({\bf x}_n, t_n)\ldots\delta J({\bf x}_1, t_1)}\right|_{J=0} \; .
\end{equation} 

This formalism was early introduced by several authors  independently,  and it is currently known as the Martin-Siggia-Rose-Janssen-DeDominicis  formalism  (MSRJD)~\cite{MSR1973,Janssen-1976,deDominicis-1976}.  More recently,  the functional formalism was generalized to deal with multiplicative noise in such a way that it is possible to treat different stochastic prescriptions in a unified way~\cite{arenas2010,Arenas2012,Arenas2012-2, Miguel2015}.  The generating functional can be represented in terms of a functional integral over four fields: two usual scalar fields  $\phi({\bf x}, t)$ and $\varphi({\bf x}, t)$ representing the order parameter and the response field respectively,  and two anticommuting Grassmann fields $\bar\xi({\bf x}, t)$ and $\xi({\bf x}, t)$.  It has the general form,
\begin{equation}
Z[J_\phi,J_\varphi]=\int {\cal D}\phi{\cal D}\varphi {\cal D}\bar\xi{\cal D}\xi\; e^{-S[\phi,\varphi,\bar\xi,\xi]+\int d^dxdt \{J_{\phi}\phi+J_{\varphi}\varphi\}} \; ,
\label{eq:ZJ}
\end{equation}
where the  ``action'' is given by~\cite{Arenas2012-2}, 
\begin{eqnarray}
\lefteqn{
S=\int d^dx dt\;  i\varphi \left\{\frac{\partial\phi}{\partial t}+\frac{\Gamma}{2} G^2 \frac{\delta H}{\delta \phi}
+\Gamma G\frac{\delta G}{\delta\phi}\bar\xi \xi\right\}+\frac{\Gamma}{2} G^2\varphi^2}
\nonumber \\
&-\int d^dx dtd^dx' dt'\;  \bar\xi({\bf x},t)K({\bf x}-{\bf x}', t-t')\xi({\bf x},t) \; , 
\label{eq:S}
\end{eqnarray}
and the kernel 
\begin{eqnarray}
K({\bf x}-{\bf x}', t-t')=\delta^d({\bf x}-{\bf x}')
\frac{d\delta(t-t')}{dt}+\frac{\Gamma}{2}\frac{\delta}{\delta\phi({\bf x},t)}\left( G^2(\phi({\bf x}',t'))\frac{ \delta H}{\delta\phi({\bf x}',t')}\right) \; .
\label{eq:K}
\end{eqnarray}
$J_\phi$ and $J_\varphi$ are sources introduced for computing different correlation functions upon derivation. For simplicity we have fixed $\beta=1$ in preceding  equations. Whether  necessary,  we could easily  restore the value of the noise intensity. 

This formalism was introduced to deal with any stochastic prescription in the same footing.  This is particularly important in the study of fluctuation theorems that  involve time reversal transformations~\cite{Arenas2012-2, Miguel2015}. A specific stochastic prescription reappears upon integration over the Grassmann fields.  Indeed,   
the retarded Green function of the  kernel $K({\bf x},t)$  and the stochastic prescription $\alpha$ are related 
by~\cite{Miguel2015}
\begin{equation}
K_R^{-1}(0,0)\sim \langle\bar\xi({\bf x}, t)\xi({\bf x}, t)\rangle\sim\alpha.
\label{eq:Kalpha}
\end{equation}

%%%%%%%%%%%%%%%%%%
\subsection{Linear response and the Fluctuation Dissipation Theorem}
\label{subsec:LR}
%%%%%%%%%%%%%%%%%%
To study linear response, we slightly perturb the system out of equilibrium  
\begin{equation}
H[\phi]\to H[\phi] -\int  d^d x \; h({\bf x}, t) \phi({\bf x}, t), 
 \label{eq:perturb}
\end{equation}
where $h({\bf x}, t)$ is a weak time dependent external field, 
and compute the 
dynamic susceptibility 
\begin{equation}
 \chi({\bf x}, {\bf x}', t,t') \equiv\left.\frac{\delta \langle \phi({\bf x}, t)\rangle_h}{\delta h({\bf x}', t')}\right|_{h=0}.
 \label{eq:chi}
 \end{equation}
By replacing Eq.  (\ref{eq:perturb}) into Eq.  (\ref{eq:S}) and computing Eq.  (\ref{eq:chi}) we find 
\begin{equation}
 \chi({\bf x}, {\bf x}', t,t') =i \frac{\Gamma}{2}\left\langle \phi({\bf x}, t) G^2(\phi({\bf x}'))\varphi({\bf x}',t') \right\rangle -\frac{\Gamma}{2} \left\langle \phi({\bf x},t)\frac{\delta G^2(\phi)}{\delta\phi({\bf x}',t')}\xi({\bf x}',t')\bar\xi({\bf x}',t') \right\rangle\; , 
 \label{eq:response}
 \end{equation}
 where $\langle\ldots\rangle$ means expectation values computed with the action of Eq. (\ref{eq:S}) with $h({\bf x}, t)=0$. 

In the case of additive noise, where $G(\phi)=1$,  the response function has the simpler form, 
\begin{equation}
\chi({\bf x}, {\bf x}', t,t') =\frac{i\Gamma}{2}\langle \phi({\bf x}, t)\varphi({\bf x}',t') \rangle \; .
\label{eq:response-Add}
\end{equation} 
For this reason, the auxiliary field $ \varphi({\bf x},t)$
is usually called the response field.  However, for multiplicative  noise, the response is more complex since 
involves, not only correlations functions of $G(\phi)$, but also contributions from
the Grassmann   sector of the model.  This is a direct consequence of the multiplicative character of the dynamics since  a variation of the ``external'' potential modifies the fluctuation properties of the system. 

Interestingly,   in the case of multiplicative systems, it is possible to write the susceptibility in a similar way to an additive system, 
\begin{equation}
\chi({\bf x}, {\bf x}', t,t') =\frac{i\Gamma}{2}\langle \phi({\bf x}, t)\tilde \varphi({\bf x}',t') \rangle \; ,
\label{eq:response-tildevarphi}
\end{equation} 
where,  from equation~(\ref{eq:response}), it is immediate to identify a {\em ``natural
response variable''}
\begin{equation}
	\tilde\varphi({\bf x}, t)= G^2(\phi)\left(\varphi({\bf x}, t)+i \frac{d\ln G^2}{d\phi}\bar\xi({\bf x}, t)\xi({\bf x}, t)\right)\; .
\label{eq:NaturalResponseVariable}
\end{equation}
For additive processes ($G=1$),   $\tilde \varphi=\varphi$. However,  for multiplicative processes, the structure of the response variable is more involved.  Indeed,   it has been shown that the response variable $\tilde\varphi$ of Eq. (\ref{eq:NaturalResponseVariable}) has a key role in the supersymmetric description of the  model~\cite{Arenas2012, Arenas2012-2}.

One of the important dynamical properties of equilibrium (either probabilistic or thermodynamic) is the ``Fluctuation Dissipation Theorem" (FDT).  It states that,   provided the system reaches equilibrium at long times,   the response function and the two-point correlation of the order parameter are not independent.  Indeed, they are given by
\begin{equation}
  \frac{1}{2} (\partial_t - \partial_{t'}) G ({\bf x}-{\bf x}', t-t') = \frac{1}{2} {\chi(|{\bf x}-{\bf x}'|, t'-t) - \chi(|{\bf x}-{\bf x}'|,t-t')},
  \label{eq:FDT1}
\end{equation}
where we have used  time translations invariance of the correlation function 
\begin{equation}
G ({\bf x}-{\bf x}', t-t')\equiv \langle \phi({\bf x}, t) \phi({\bf x},' t')\rangle\; .
\end{equation}
Equation (\ref{eq:FDT1})  can be re-written using  causality  as
\begin{equation}
  \chi({\bf x}-{\bf x}', t-t') = - (\partial_t - \partial_{t'}) G ({\bf x}-{\bf x}', t-t') \Theta(t-t'),
  \label{eq:FDT2}
\end{equation}
where $\Theta(t)$ is the Heaviside step function.  Equation (\ref{eq:FDT2})  is the traditional form of the FDT.  Perhaps, a more usual form,  useful to interpret experimental or numerical simulation data,   is given by integrating over ${\bf x}$ and expressing  correlations and responses  in terms of the Fourier transform at zero wave vector, 
\begin{equation}
	\tilde G(\omega) = \frac{\mathcal{I}m\ \tilde\chi(\omega)}{\omega}. 
 \label{eq:FDT-FT}
\end{equation}

The general form of the FDT depends only on equilibrium properties and does not depend on whether the dynamical process is additive or multiplicative.   However, the explicit form of the susceptibility does depend. Indeed,   the response function is given by Eq. (\ref{eq:response}), where the influence of the multiplicative dynamics is evident. 
The constraints imposed on correlation functions by the FDT are exact.  Then, they are a good arena to test different type of approximations.   In fact, we will use it to check the consistency of the $\epsilon$-expansion of the Dynamical Renormalization Group equations.  
It is very interesting to note that extensions of the
FDT  for out-of-equilibrium steady states can be also formulated~\cite{Marconi2008,ArBaCuZoGus2016}.  

 %%%%%%%%%%%%%%%%%%%%%%%%%%%%%%%%%%%%%%%
\subsection{Real scalar field with $Z_2$ symmetry}
\label{subsec:Model}
In this subsection we  present a specific model. We show the structure of the action of Eq. (\ref{eq:S}) in the case of a scalar model with $Z_2$ symmetry. 
We begin by considering the quartic Hamiltonian of  a real scalar field $\phi({\bf x})$: 
\begin{equation}
H=\int_{\frac{1}{\Lambda}} d^d x \left\{\frac{1}{2}\left| {\bf \nabla}\phi\right|^2+\frac{r}{2}\phi^2+\frac{u}{4!}\phi^4\right\}\; , 
\label{eq:Hamiltonian}
\end{equation}
where $\Lambda$ is an ultraviolet momentum cut-off.  $\{r,u\}$ are the quadratic and quartic coupling constants respectively.    In addition,  we consider a general diffusion function with the same symmetry of the Hamiltonian,  $G(\phi)=G(-\phi)$.  Without loosing generality,  we can choose a diffusion function satisfying $G(0)=1$, which is the usual additive noise value. Then, we can write
\begin{equation}
G^2(\phi)=1+\sum_{n=1}^{\infty} g_n \phi^{2n}  \; , 
\label{eq:G2}
\end{equation}
where $g_n$, with $n=1,2,\ldots$, are coupling constants defining the multiplicative noise distribution.

By replacing Eqs. (\ref{eq:Hamiltonian}) and (\ref{eq:G2}) into Eq. (\ref{eq:S}),  we obtain the effective action for this model. This action includes terms of all orders in powers of  the $\phi$ fields. However, since we are interested in the dynamics near the phase transition, it is possible to simplify this action. 

Related with the static properties, this system has an upper critical dimension $d_c=4$. For $d>4$,  $u$ is an irrelevant scaling variable  and  the critical point describing the phase transition is a Gaussian one.  In this sense,  all the other higher order couplings are also irrelevant.   For dimensions very near the upper critical one,   $d = 4-\epsilon$ with $\epsilon\ll 1$,   $u$ turns out to be relevant and the RG flux goes to the very well known Wilson-Fisher fixed point~\cite{Wilson-1972}.   Higher order  couplings remain irrelevant related to the new fixed point, for small values of $\epsilon$.  For this reason, in order to study the phase transition near the upper critical dimension, the quartic Hamiltonian is a good starting point, {\em i.e.}, we can safely ignore couplings of the order $\phi^6$ and greater.  

There are some terms in Eq. (\ref{eq:S}) that deserve special attention since contain contributions from the multiplicative dynamics.  One of them is
\begin{eqnarray}
G^2 \frac{\delta H}{\delta \phi}&=& -\nabla^2\phi+r\phi+ \left(\frac{u}{3!}+r g_1\right)\phi^3+
\left(\frac{g_1 u}{3!}+r g_2\right)\phi^5 +\left(g_1+g_2\phi^2 \right)\phi^2\nabla^2\phi +O(\phi^7) \; .
\end{eqnarray}
In order to study criticality in $d=4-\epsilon$,  it is sufficient to retain only cubic terms in this expression, 
\begin{eqnarray}
G^2 \frac{\delta H}{\delta \phi}&\sim -\nabla^2\phi+r\phi+ \frac{u}{3!}\phi^3+ O(\phi^5) \; , 
\end{eqnarray}
where we have shifted the coupling constant $u\to u-6 r g_1$(we will show that this shift is relevant only at order $\epsilon^3$).   We have also discarded cubic terms with derivatives, since these terms will be  irrelevant.  

Another interesting  term, coming from the multiplicative structure,  is
\begin{eqnarray}
 G\frac{\delta G}{\delta\phi}=\frac{1}{2} \frac{\delta G^2}{\delta\phi}=\sum_n n g_n \phi^{2n-1}\; .
 \label{eq:GdeltaG}
\end{eqnarray}
With the same  spirit, we retain in this expression just the first term, 
\begin{eqnarray}
 G\frac{\delta G}{\delta\phi}\sim  g_1 \phi +O(\phi^3).
\end{eqnarray}
In the following sections,  after a detailed analysis of the DRG transformations,  we will analyze  the effect of  other 
couplings $g_n$, with $n\ge 2$. We can advance at this point that it is sufficient to consider just $g_1$, since $g_2,g_3,\ldots$ will be irrelevant, at least at order $\epsilon^2$.    

By substituting the form of $G$ and $H$ in Eq. (\ref{eq:S}) and making the above mentioned approximations,  we find the effective action of the model that can be split  as 
\begin{equation}
S=S_0+S_I.   
\end{equation}
 The quadratic part, $S_0$, is given by
\begin{equation}
S_0=\int_{\frac{1}{\Lambda}} d^dx dt\; \left\{ i\varphi \Delta \phi + \frac{\Gamma}{2} \varphi^2-\bar\xi \Delta \xi\right\}
\label{eq:S0}
\end{equation}
where the differential operator 
\begin{equation}
\Delta=\partial_t+\frac{\Gamma}{2} \left(-\nabla^2+r\right).
\end{equation}
The interacting part of the action reads, 
\begin{eqnarray}
S_I=\int_{\frac{1}{\Lambda}} d^dx dt\; 
\left\{ \frac{iu\Gamma}{3! 2}\varphi \phi^3-\frac{u\Gamma}{4} \phi^2\bar\xi\xi
+\frac{g\Gamma}{2} \varphi^2\phi^2+ig\Gamma\varphi\phi\bar\xi\xi\right\} \; .
\label{eq:SI} 
\end{eqnarray}
where we are calling $g\equiv g_1$ to simplify notation. 
The first two terms are the usual ones describing an additive noise overdamped dynamics. The last two terms codify the effect of multiplicative noise dynamics.  Indeed,   the coupling of the response field $\varphi$ with the Grassmann variables $\bar\xi\xi$ is a clear distinguishing characteristic of multiplicative noise.   In the next section, we explicitly build the DRG equations for this model.

%%%%%%%%%%%%%%%%%%%%%%
\section{ Dynamical Renormalization Group} 
\label{Sec:DRG}
%%%%%%%%%%%%%%%%%%%%%%%
In order to study dynamics of the order parameter very near the critical point,  we apply  the Wilson renormalization group program  to the action of Eqs. (\ref{eq:S0})-(\ref{eq:SI}). This procedure is known as Dynamical Renormalization Group (DRG). 

%%%%%%%%%
\subsection{ DRG Transformation}
\label{subsec:DRGT}

Following the usual Wilson procedure, in order to build a DRG transformation we firstly reduce the ultraviolet cut-off to $\Lambda/b$, with $b>1$ and split the fields:
 \begin{eqnarray}
\phi({\bf x}, t)&=&\phi^<({\bf x}, t)+\phi^>({\bf x}, t) \; ,\\
\varphi({\bf x}, t)&=&\varphi^<({\bf x}, t)+\varphi^>({\bf x}, t)\; , \\
\bar\xi({\bf x}, t)&=&\bar\xi^<({\bf x}, t)+\bar\xi^>({\bf x}, t)\; , \\
\xi({\bf x}, t)&=&\xi^<({\bf x}, t)+\xi^>({\bf x}, t)\; , 
\end{eqnarray}
in such a way that the Fourier transformed  fields with superscript ``$<$"  have support within the sphere $k<\Lambda/b$, while the fields labeled by ``$>$" have support on a spherical shell $\Lambda/b<k<\Lambda$. In these expressions, $k=|{\bf k}|$.
For instance, for the field $\phi$, we have 
\begin{eqnarray}
\phi^<({\bf x}, t)&=&\int \frac{d^dk}{(2\pi)^d} \frac{d\omega}{2\pi}   \Theta\left(\frac{\Lambda}{b}- k\right)\; \phi({\bf k}, \omega) e^{-i(\omega t- {\bf k}\cdot {\bf x})}\; , 
\label{eq:FourierPhiM} \\
\phi^>({\bf x}, t)&=&\int \frac{d^dk}{(2\pi)^d} \frac{d\omega}{2\pi}   
\Theta\left(k-\frac{\Lambda}{b}\right)\Theta\left(\Lambda- k\right)\; \phi({\bf k}, \omega) e^{-i(\omega t- {\bf k}\cdot {\bf x})} \; , 
\label{eq:FourierPhim}
\end{eqnarray}
where $\Theta(k)$ is the usual Heaviside step distribution.  The same definitions are used for the other fields.

 We define the transformed action $S'_{\Lambda/b}[\phi^<,\varphi^<,\bar\xi^<,\xi^<]$, depending only on the field component with lower momentum,  by integrating out the fields with momentum higher than $\Lambda/b$ in the following way
\begin{equation}
e^{-S'_{\Lambda/b}[\phi^<,\varphi^<,\bar\xi^<,\xi^<]}=\int {\cal D}\phi^>{\cal D}\varphi^> {\cal D}\bar\xi^>{\cal D}\xi^>\; e^{-S_\Lambda[\phi,\varphi,\bar\xi,\xi]} \; .
\label{eq:DRGT}
\end{equation}
Since the transformed action, $S'_{\Lambda/b}$,  has a momentum cut-off $\Lambda/b$,  in order to compare it with the original one, $S_{\Lambda}$,  with cut-off $\Lambda$,  we re-scale momentum and frequency as 
\begin{eqnarray}
k'&=& b \; k  \; ,
\label{eq:bk} \\
\omega' &=&  b^z \omega \; ,
\label{eq:bzomega}
\end{eqnarray}
where $z$ is the dynamical critical exponent.
We also need to re-scale the fields in the following way, 
\begin{eqnarray}
\varphi^<(k'/b, \omega'/b^z)&=& b^{\frac{d-2+2z-\eta_1}{2}}\varphi(k', \omega') \; , 
\label{eq:varphiscaled}\\
\phi^<(k'/b, \omega'/b^z)&=& b^{\frac{d+2+2z+\eta_2}{2}}\phi(k', \omega') \; ,
\label{eq:phiscaled}\\
\bar\xi^<(k'/b, \omega'/b^z)&= &b^{\frac{d+2z+\eta_3}{2}}\bar\xi(k', \omega') \; ,
\label{eq:barxiscaled}\\
\xi^<(k'/b, \omega'/b^z)&=& b^{\frac{d+2z+\eta_3}{2}}\xi(k', \omega')\; ,
\label{eq:xiscaled}
\end{eqnarray}
where $\eta_1,\eta_2,\eta_3$ are anomalous dimensions for the fields $\varphi$, $\phi$ and $\{\bar \xi, \xi\}$ respectively.  
 
The transformation of Eq. (\ref{eq:DRGT}), together with the re-scalings from Eq. (\ref{eq:bk}) to Eq. (\ref{eq:xiscaled}) 
define a Dynamical Renormalization Group Transformation (DRGT).  After this transformation, the transformed action $S'$ is defined in the same momentum domain of $S$ and can be compared.   By successive application of this transformation we  generate a DRG flux of the actions or equivalently of its coupling constants.  By considering an infinitesimal transformation $b=1+\delta\ell$, with $\delta\ell<<1$, we find DRG differential equations for $\{r(\ell),u(\ell),g(\ell), \Gamma(\ell)\}$.

Of course, the main difficulty of  this procedure is to compute the functional integral of Eq. (\ref{eq:DRGT}). To do this, we implement a perturbative calculation. To second order in the couplings  we find,  
\begin{equation}
S'_{\Lambda/b}=  S^<+ \langle S_I\rangle_{S_0^>}-\frac{1}{2} \left(\langle S_I^2\rangle_{S_0^>}-\langle S_I\rangle_{S_0^>}^2\right) \; , 
\label{eq:Sperturbative}
\end{equation}
where $S^<=S_0^<+S_I^<$ means the action of Eqs. (\ref{eq:S0}) and (\ref{eq:SI}) with the substitution of the fields $\phi,\varphi,\bar\xi,\xi$ by $\phi^<,\varphi^<,\bar\xi^<,\xi^<$.  The notation $\langle\ldots\rangle_{S_0^>}$ means the mean value computed with the quadratic action $S_0$ of Eq. (\ref{eq:S0}) but with the momentum constrained to the small shell  $\Lambda/b<k<\Lambda$. 

The first term of Eq. (\ref{eq:Sperturbative}) is the so called tree level approximation. Applying an infinitesimal DRGT at this level of approximation we immediately obtain the flux equations, 
\begin{eqnarray} 
\frac{dr}{d\ell} &=&2r  \; ,
\label{eq:rtree} \\
	\frac{du}{d\ell}&=&(4-d) u \; , 
	\label{eq:utree}
	\\
	\frac{dg}{d\ell}& =& \left(2-d\right) g \; ,
	\label{eq:gtree}
	\\
	\frac{d\Gamma}{d\ell}&=&\left(z-2\right) \Gamma  \; .
	\label{eq:gammatree}
\end{eqnarray}
%%%%%%%%%%%%%%
Equations (\ref{eq:rtree}) and (\ref{eq:utree}), for $r$ and $u$,  are the usual ones of the equilibrium theory. The last two equations (Eqs. (\ref{eq:gtree}) and (\ref{eq:gammatree})),  are proper of the dynamics.  The only fixed point is the Gaussian one,  $r^*=u^ *=g^*=0$.   As it is well known, $r$ has scaling dimension 2, meaning that it is a relevant scaling variable.  On the other hand, the scaling dimension of $u$ is $4-d$; this defines the upper critical dimension $d_c=4$.  Above $d_c$,  the coupling $u$ is irrelevant, meaning that the Gaussian fixed point is stable and rules the phase transition. For dimensions smaller than the critical one, the coupling $u$ is relevant, and the flux goes away from the Gaussian fixed point.   From equation (\ref{eq:gammatree}), we can read the dynamical critical exponent $z=2$,  adjusted to make the diffusion coefficient $\Gamma$ marginal. Eq.  (\ref{eq:gtree}),  controls the multiplicative noise coupling $g$.  We observe that it has scaling dimension $2-d$. At this level of approximation, $g$ is an irrelevant coupling for any dimension  $d>2$.  At exactly $d=2$,  $g$ is a marginal scaling variable.  Therefore, at tree level, the effective dynamics is controlled by a usual additive Langevin equation.  
 
 %%%%%%%%%%%%%%%%%%%%%%%%%%%%
\subsection{$\epsilon$-expansion}
\label{subsec:epsilon}
 For $d<4$,  $u$ is a relevant scaling variable,  the Gaussian fixed point is unstable and we need to compute fluctuations to determine the flux diagram; they are given by the second and third terms of r.h.s of Eq. (\ref{eq:Sperturbative}). 
The static result is very well known.  Indeed, for $d= 4-\epsilon$ with $\epsilon<<1$,  the flux runs to the celebrated Wilson-Fisher fixed point~\cite{Wilson-1972}, that can be systematically computed through an expansion in powers of $\epsilon$.  Here we are interested in the dynamics near this nontrivial fixed point. 

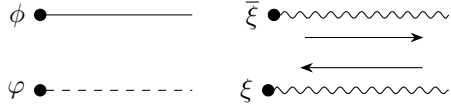
\begin{figure}
\begin{center}
\begin{tikzpicture}
	\begin{feynman}
		%DEFINE TODOS OS VERTICES	
		\node[dot, label= left:\( \phi\)](a1);
		\vertex[right=2cm of a1,] (a2);
		
		\node[below=1cm of a1,dot,label=left:\(\varphi\)] (b1);
		\vertex[right=2cm of b1] (b2);

		\node[right=1cm of a2,dot,label=left:\(\overline\xi\)] (c1);
		\vertex[right=2.3cm of c1] (c2);

		\vertex[below=1cm of c2] (d1);
		\node[left=2.3cm of d1,dot,label=left:\(\xi\)] (d2);
		
		%DESENHA OS DIAGRAMAS
		\diagram*[small]{
			(a1) -- [plain] (a2),
			(b1) -- [scalar] (b2),
			(c1) -- [boson,  momentum'=\(\)] (c2),
			(d1) -- [boson,  momentum'=\(\)] (d2)};
	\end{feynman}
\end{tikzpicture}
\end{center}
\caption{Feynman rules conventions.  External lines representing  the fields $\phi^<, \varphi^<, \bar\xi^<, \xi^<$.  We assign a solid external  line to represent the order parameter field $\phi$ and the external dashed line to represent the response field $\varphi$.  The Grassmann fields are represented by wiggled lines.  The arrows in the Grassmann lines are used to distinguish $\bar\xi$ from $\xi$.  The arrow points out of the field $\bar\xi$ and points into the field $\xi$.  }
\label{fig:lines}
\end{figure}

 The most efficient way to organize the expansion is by using Feynman diagrams and grouping terms by the number of internal loops. We build the Feynman diagrams by defining four lines; one for each of the four fields as shown in Figure~\ref{fig:lines}. 
%%%%%%%%%%%%%%
%%%%%%%%%%%%%%
On the other hand, we have four types of correlation functions depicted as internal lines in Figure~\ref{fig:propagators}. 
%%%%%%%%%%%%%%
\begin{figure}
\begin{center}
\begin{tikzpicture}
	\begin{feynman}
		%DEFINE TODOS OS VERTICES	
		\vertex (a0){\( \langle\phi(\textbf{x},t)\phi(\textbf{x}',t')\rangle = \)};
		\node[right=of a0,dot] (a1);
		\node[right=2cm of a1,dot] (a2);

		\vertex [below=1cm of a0](b0){\( \langle\phi(\textbf{x},t)\varphi(\textbf{x}',t')\rangle = \)};
		\node[right=of b0,dot] (b1);
		\node[right=2cm of b1,dot] (b2);
		
		\vertex [below=1cm of b0](c0){\( \langle\overline\xi(\textbf{x},t)\xi(\textbf{x}',t')\rangle = \)};
		\node[right=of c0,dot] (c1);
		\node[right=2cm of c1,dot] (c2);		
		
		\vertex [below=1cm of c0](d0){\( \langle\xi(\textbf{x},t)\overline\xi(\textbf{x}',t')\rangle = \)};
		\node[right=of d0,dot] (d1);
		\node[right=2cm of d1,dot] (d2);

		%DESENHA OS DIAGRAMAS
		\diagram*{
			(a0) -- [plain](a1) -- [plain] (a2),
			(b0) -- [plain](b1) -- [fermion] (b2),
			(c1)--[charged boson](c2),	
			(d1)--[anti charged boson](d2)
		};

	\end{feynman}
\end{tikzpicture}
\end{center}
\caption{Graphic representation of correlation functions.   The continuous line connecting  two points represents $G^>({\bf x}-{\bf x}',t-t')=\langle \phi({\bf x},t)\phi({\bf x}',t')\rangle_{S_0^>}$.   This correlation function is symmetric in space and time.   We represent the response function  $R^>({\bf x}-{\bf x}',t-t')=\langle \phi({\bf x},t)\varphi({\bf x}',t')\rangle_{s_0^>}\Theta(t-t')$ with a continuous line with an arrow pointing  in the direction of decreasing time.   Notice that the retarded response function is not symmetric in time.  For this reason, it is necessary the arrow to indicate the time direction.   The Grassmann fields propagators are indicated by wiggled lines with an arrow pointing in the direction of the field $\xi$.  Since there are no response propagator 
$\langle\varphi\varphi\rangle=0$,  there is no dashed internal lines in the Feynman diagrams.  }
\label{fig:propagators}
\end{figure}
%%%%%%%%%%%%%%%%%

From $S_0$, given by Eq. (\ref{eq:S0}),  we can read two propagators: 
\begin{equation}
G^>({\bf x}-{\bf x}',t-t')=\langle \phi({\bf x},t)\phi({\bf x}',t')\rangle_{S_0^>} \; , 
\end{equation}
represented by the first internal continuous line of Figure~\ref{fig:propagators},
and 
\begin{equation}
R^>({\bf x}-{\bf x}',t-t')=\langle \phi({\bf x},t)\varphi({\bf x}',t')\rangle_{S_0^>}\Theta(t-t')\; , 
\label{eq:Ralpha}
\end{equation} 
represented by an internal continuous line with an arrow in the same figure.   The arrow goes in the direction of decreasing time,  {\em i.e.},  from $t\to t'$ in  Eq.  (\ref{eq:Ralpha}).

The Grassmann field propagator is related with the response function
 \begin{equation}
 \langle \bar\xi({\bf x},t)\xi({\bf x}',t')\rangle=-i R^>({\bf x}-{\bf x}',t-t') \label{eq:grassmannpropagator}
 \end{equation} 
 and is represented by an internal wiggled  line in Figure~\ref{fig:propagators}.  
 The bare response fields are not correlated $\langle \varphi({\bf x},t)\varphi({\bf x}',t')\rangle=0$; for this reason there are no internal dashed  lines in the Feynman diagrams. 
  
 Explicitly computing the propagators in Fourier space we find,   
\begin{eqnarray}
\tilde G^>({\bf k},\omega)&=&
\frac{\Gamma}{\omega^2+\frac{ \Gamma^2}{4} \left(k^2+r\right)^2}\; ,
\label{eq:G+}
\\
\tilde R^>({\bf k},\omega)&=&
\frac{-i}{-i\omega+ \frac{\Gamma}{2} \left(k^2+r\right)}\; .
\label{eq:R+}
\end{eqnarray}
It is important to remember that the  superscript ``$>$"  means that the momentum $k$ is constrained to a thin spherical shell  $(1-\delta\ell)<k/\Lambda <1$.

\begin{figure}
	\begin{center}
		\includegraphics[width=0.5\textwidth]{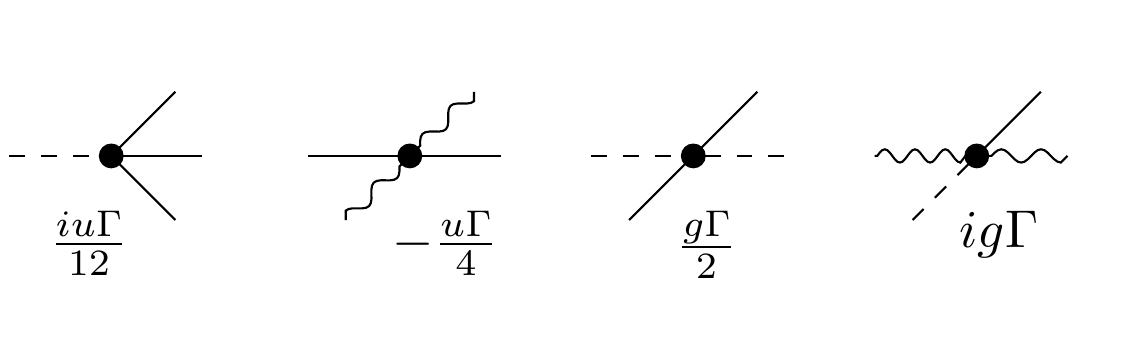}
	\end{center}
	\caption{Vertices as read from the  action $S_I$ of Eq. (\ref{eq:SI}).  The dashed line represents the field $\varphi({\bf x}, t)$ while the solid line represents the field  $\phi({\bf x}, t)$.  The wiggled lines correspond to the Grassmann fields $\{\bar\xi({\bf x}, t),
	\xi({\bf x}, t)\}$.  The first two vertices are the usual ones in the functional representation of an additive Langevin equation,  while the last two vertices,  proportional to $g$,   arise from the multiplicative noise distribution. }
	\label{fig:vertex}
\end{figure}

The Feynman rules are completed by  four vertices  read from equation (\ref{eq:SI}) and depicted in Figure~\ref{fig:vertex}. 
%%%%%%%%%%%%%%%%%%%%%%%%%%%%%
%%%%%%%%%%%%%%%%%%%
The first two vertices, proportional to $u$,  are the usual ones in an additive stochastic process.  The last two vertices, proportional to $g$, come from the multiplicative noise distribution.  Indeed, the coupling of the response field $\varphi$ with the Grassmann fields $\bar\xi, \xi$ is a hallmark of multiplicative noise~\cite{Arenas2012-2}.

\begin{figure}
	\begin{center}
\includegraphics[width=0.5\textwidth]{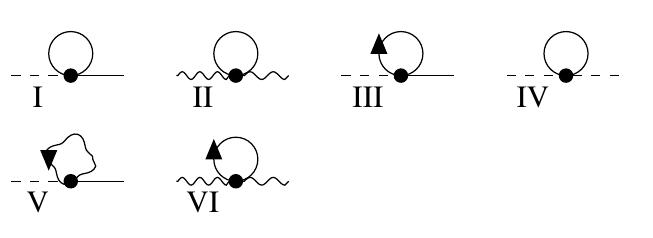}
	\end{center}
	\caption{First order corrections.  Internal continuous lines represent the $G^>$ propagator while internal arrowed lines correspond to  the response function $R^>$.   Wiggled internal lines represent Grassmann propagators.  The diagrams I, II, III, V and VI are contributions that essentially renormalize the $r$ parameter.  Diagram IV  contributes to renormalize $\Gamma$ and it appears in the anomalous dimensions $\eta_1$, $\eta_2$ . }
	\label{fig:1V1loop}
\end{figure}

\begin{figure}
	\begin{center}
\includegraphics[width=0.45\textwidth]{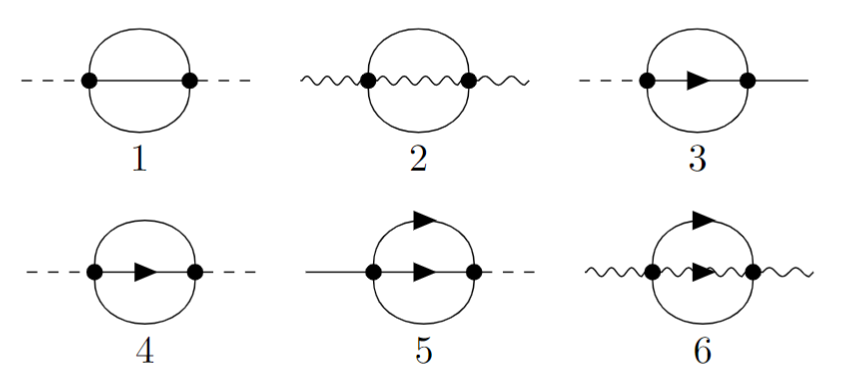}
	\end{center}
	\caption{Two loop second order diagrams.  The line conventions are the same of Figure~\ref{fig:1V1loop}. 
	Diagrams 1, 2 and 3 are of order $u^2$ while diagrams 4, 5 and 6 are of order $ug$, coming from the multiplicative noise vertices.  All these diagrams,  in addition to the contribution to the renormalization of $r$ and $\Gamma$, contribute to the anomalous dimensions $\eta_1, \eta_2, \eta_3$.}
	\label{fig:2V2loop}
\end{figure}

\begin{figure}
	\begin{center}
		\includegraphics[width=0.45\textwidth]{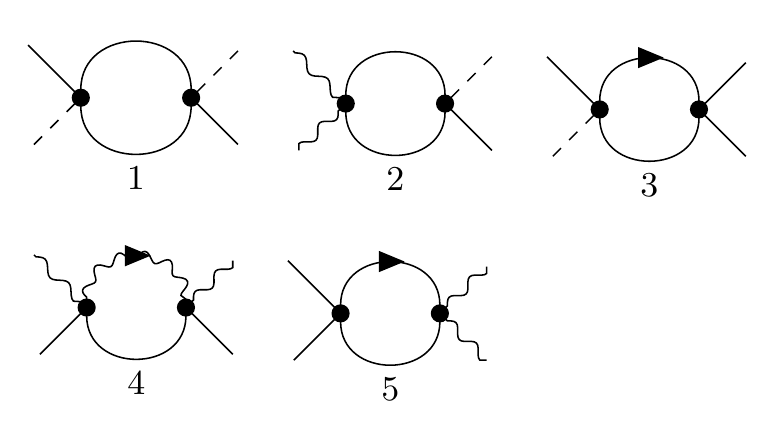}
	\end{center}
	\caption{One-loop vertex corrections proportional to $u^2$. The line conventions are the same of Figure~\ref{fig:1V1loop}.    The first diagram, proportional to $\phi^2\varphi^2$, and the second one,  proportional to 
	 $\varphi\phi\bar\xi\xi$,  correct the multiplicative constant $g$.   They are the main contributions that show how nonlinear terms in the Hamiltonian,  proportional to $u$,  generate dynamic multiplicative couplings.}
	\label{fig:u2_1loop}
\end{figure}

\begin{figure}
	\begin{center}
		\includegraphics[width=0.45\textwidth]{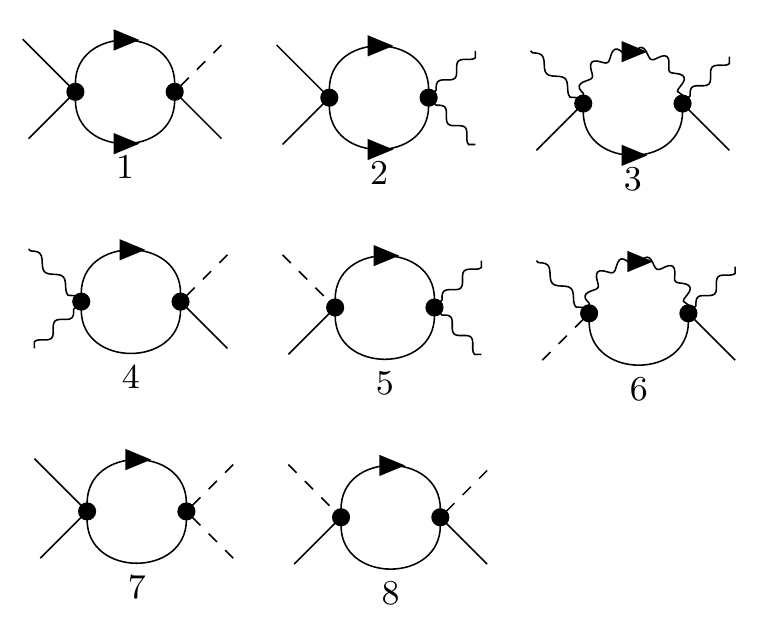}
	\end{center}
		\caption{One-loop vertex corrections proportional to $ug$.  The line conventions are the same of Figure~\ref{fig:1V1loop}. }
	\label{fig:ug_1loop}
\end{figure}

\begin{figure}
	\begin{center}
		\includegraphics[width=0.35\textwidth]{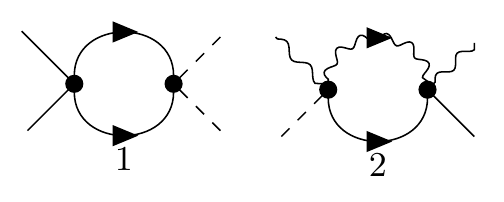}
	\end{center}
\caption{One-loop vertex corrections proportional to $g^2$. The line conventions are the same of Figure~\ref{fig:1V1loop}. }
\label{fig:g2_1loop}
\end{figure}

First order corrections are given by $\langle S_I\rangle_{S_0^>}$ (the second term of the r.h.s of Eq. (\ref{eq:Sperturbative})).  The corresponding single vertex diagrams are depicted in  Figure~\ref{fig:1V1loop}.
%%%%%%%%%%%%%%%%%%%
%%%%%%%%%%%%%%%%%%%
All these diagrams are constants and contribute to renormalize $r$ and $\Gamma$.  Indeed,  diagrams I,  III and V renormalize the term $r\varphi\phi$.  The diagrams II and VI renormalize $r\bar\xi\xi$ while diagram IV renormalizes $\Gamma \varphi^2$.   The couplings $r$ as well as $\Gamma$ also have second order corrections coming from the last term of Eq. (\ref{eq:Sperturbative}).  They are given by the two-loop diagrams in Figure~\ref{fig:2V2loop}.
%%%%%%%%%%%%%%%%%%%%%%
%%%%%%%%%%%%%%%%%%%%%%
The  four vertices proportional to $u$ and $g$,  shown in Figure~\ref{fig:vertex},  are corrected by second order one-loop diagrams proportional to $u^2$,  $ug$ and $g^2$.  We depicted these contributions in Figures~\ref{fig:u2_1loop}, \ref{fig:ug_1loop} and~\ref{fig:g2_1loop}, respectively. 
%%%%%%%%%%%%%%%%%%%%%%
%%%%%%%%%%%%%%%%%%%%%%	
%%%%%%%%%%%%%%%%%%%%%%
%%%%%%%%%%%%%%%%%%%%%%
%%%%%%%%%%%%%%%%%%%%%%
%%%%%%%%%%%%%%%%%%%%%%

We computed all the diagrams and, after re-scaling momentum, frequency and fields, we built the DRG equations. 
We show this calculation in detail in Appendix \ref{App:DRG-equations},  \ref{App:Integrals} and \ref{App:Rescaling}.  The perturbative calculation has real meaning very near the upper critical dimension, where higher order coupling terms are irrelevant and can be ignored. For this reason we have performed an $\epsilon=4-d$ expansion and we keep terms up to order $\epsilon^2$. 

The  DRG equations are  displayed  in a simpler way  by using dimensionless coupling constants. For this purpose we  have rescaled the couplings 
\begin{eqnarray}
\Lambda^{-2} \; r&\to & r  \; ,  \\
\frac{\Omega_d}{(2\pi)^d} \; \Lambda^{d-4} \;  u&\to &  u \; ,    \\
\frac{\Omega_d}{(2\pi)^d}\; \Lambda^{d-2}  \; g&\to & g \; , 
\end{eqnarray} 
where $\Omega_d$  is the area of a $(d-1)$-dimensional sphere of radius one. With this, the  new couplings $\{ r,u,g\}$ are now dimensionless. 
Keeping only the relevant terms that contribute to the fixed points to order $\epsilon^2$, we get the dimensionless DRG equations (see Appendix \ref{App:DRG-equations} , \ref{App:Integrals} and \ref{App:Rescaling} for explicit details of the calculation)
%%%%%%%%%%%%%%%
\begin{eqnarray} 
\frac{dr}{d\ell} &=&2r+\frac{u(1-r)}{2}-\frac{u^2}{6}-2 \alpha  g 
\label{eq:r}
\\
\frac{du}{d\ell}&=&u\left[\epsilon-\frac{3 u(1-2 r)}{2}-\frac{2u^2}{27}+4 g \right]
	\label{eq:u}
	\\
\frac{dg}{d\ell}& =&-2 g  +\frac{u^2}{4}
	\label{eq:g}
\\
	\frac{d\Gamma}{d\ell}&=& \Gamma  \left(z-2-\frac{ u^2}{54}\right)
	\label{eq:gamma}
\end{eqnarray}
%%%%%%%%%%%%%%
This set of equations is one of the main results of the paper. 
The multiplicative coupling constant $g$ corrects the usual RG equations for $r$ and $u$ as can be seen in the last term of Eqs. (\ref{eq:r}) and (\ref{eq:u}). It is interesting to notice that the last term of equation (\ref{eq:r}) depends on the stochastic prescription $\alpha$. As we have anticipated in Section~\ref{subsec:Equilibrium}, this fact was expected since the critical temperature (codified in $r$) depends on the prescription in multiplicative stochastic systems. 
This term comes from contributions of the tadpole diagrams III and V in Figure~\ref{fig:1V1loop}. 
The equilibrium properties of an additive process do not depend on the stochastic prescription. This is reflected  in the fact that the Bosonic and Grassmann tadpole contributions cancel each other. However, in the multiplicative noise case, tadpole contributions do not necessarily cancel out~\cite{arenas2010}. The tadpole diagrams contribute with the response function computed at the same position and time.  As noticed in Eqs. (\ref{eq:Kalpha}) and  (\ref{eq:Ralpha}), $R^>_{\rm retarded}(0,0)\sim -i \alpha$, where the parameter $\alpha$ corresponds with the stochastic prescription~\cite{Arenas2012-2}(see Appendix~\ref{App:Integrals} for explicit details of the calculation).  

Equation~\ref{eq:g} is the key equation that controls the fate of the multiplicative coupling. We observed that the leading correction to the tree level result is proportional to $u^2$, coming from the diagram 1 of Figure~\ref{fig:u2_1loop}.  This diagram occurs naturally from contracting two vertices proportional to  $ iu\Gamma $ (first vertex in Figure~\ref{fig:vertex}),     producing a correction to the  multiplicative vertex  $g\Gamma$ (third vertex in Figure~\ref{fig:vertex}).  This is the most important diagram that drives the DRG flux to a novel multiplicative fixed point.  It is also worth to note that diagrams proportional to $ ug $ and $ g^2 $ from Figures~\ref{fig:ug_1loop} and~\ref{fig:g2_1loop} respectively, will also contribute to renormalize  $g$.  However, they do not appear in the system of Eqs. (\ref{eq:r})-(\ref{eq:g}) because  they will produce contributions of order  $ \epsilon^3 $ in the fixed points. In Appendix~\ref{App:DRG-equations}, we show the complete two-loop DRG equations and the $\epsilon-$expansion approximation that leads to Eqs.  (\ref{eq:r})-(\ref{eq:gamma}).

The system of Eqs. (\ref{eq:r})-(\ref{eq:g}) has several fixed points.   The stability depends essentially on dimensionality.   In the following, we describe in detail each fixed point. 

\subsubsection{The Gaussian Fixed Point}
\label{subsub:Gaussian}
The DRG Eqs.  (\ref{eq:r})-(\ref{eq:g}) have the well known  Gaussian  fixed point given by $u_G=r_G=g_G=0$  with $z=2$.  At this point,  the  Hamiltonian is quadratic and the dissipative dynamics is additive.  

We can study its stability by linearizing Eqs.  (\ref{eq:r})-(\ref{eq:g})  around this fixed point.   We get 
 \begin{equation}
\left(
\begin{array}{c}
\frac{d r}{d\ell} \\
\frac{d u}{d\ell} \\
\frac{d g}{d\ell} \\
\end{array}
\right)
=
\left(
\begin{array}{ccc}
T_{rr} & T_{ru} & T_{rg} \\
T_{ur} & T_{uu} & T_{ug} \\
T_{gr} & T_{gu} & T_{gg}
\end{array}
\right) 
 \left(
\begin{array}{c} 
r \\
u \\
g\\
\end{array}
\right)  \;, 
\end{equation}
where the matrix is given by
\begin{equation}
	T_G=\left(
	\begin{array}{ccc}
		2  &~~\frac{1}{2} & -2 \alpha  \\
		&  &  \\
		0 &~~\epsilon  & 0 \\
		&  &  \\
		0 &~~0 & -2 \\
	\end{array}
	\right) +O\left(\epsilon^3\right) \; .
\end{equation} 
By diagonalizing $T_G$, we obtain the eigenvalues and eigenvectors that will define the relevant and irrelevant scaling variables. 
It is very simple to read the following  eigenvalues for the Gaussian fixed point:
\begin{eqnarray}
	y_r^G&=& 2 
	\label{eq:yrG}\\
	y_u^G&=& \epsilon+ O(\epsilon^3)
	\label{eq:yuG}  \\
	y_g^G&=& -2
	\label{eq:ygG}
\end{eqnarray}
and the directions of the scaling variables, given by the corresponding normalized eigenvectors,
\begin{eqnarray}
	{\bf v}_r^G&=&\left\{1,0,0\right\} 
	\label{eq:vrgG} \\
{\bf v}_u^G&=&\frac{1}{\sqrt{17}}
	\left\{-1,4, 0 \right\}+ \frac{2}{17\sqrt{17}}
	\left\{4 ,1,0  \right\}\epsilon+
	O\left(\epsilon ^2\right)
	\label{eq:vuG} \\
	{\bf v}_g^G&=&\frac{1}{\sqrt{\alpha ^2+4}}\left\{\alpha,0,2\right\}
	\label{eq:vgG}
\end{eqnarray}

\begin{figure*}[htb]
	\begin{center}
		\subfigure[]
		{\label{fig:FPur4d}
			\includegraphics[width=0.31\textwidth]{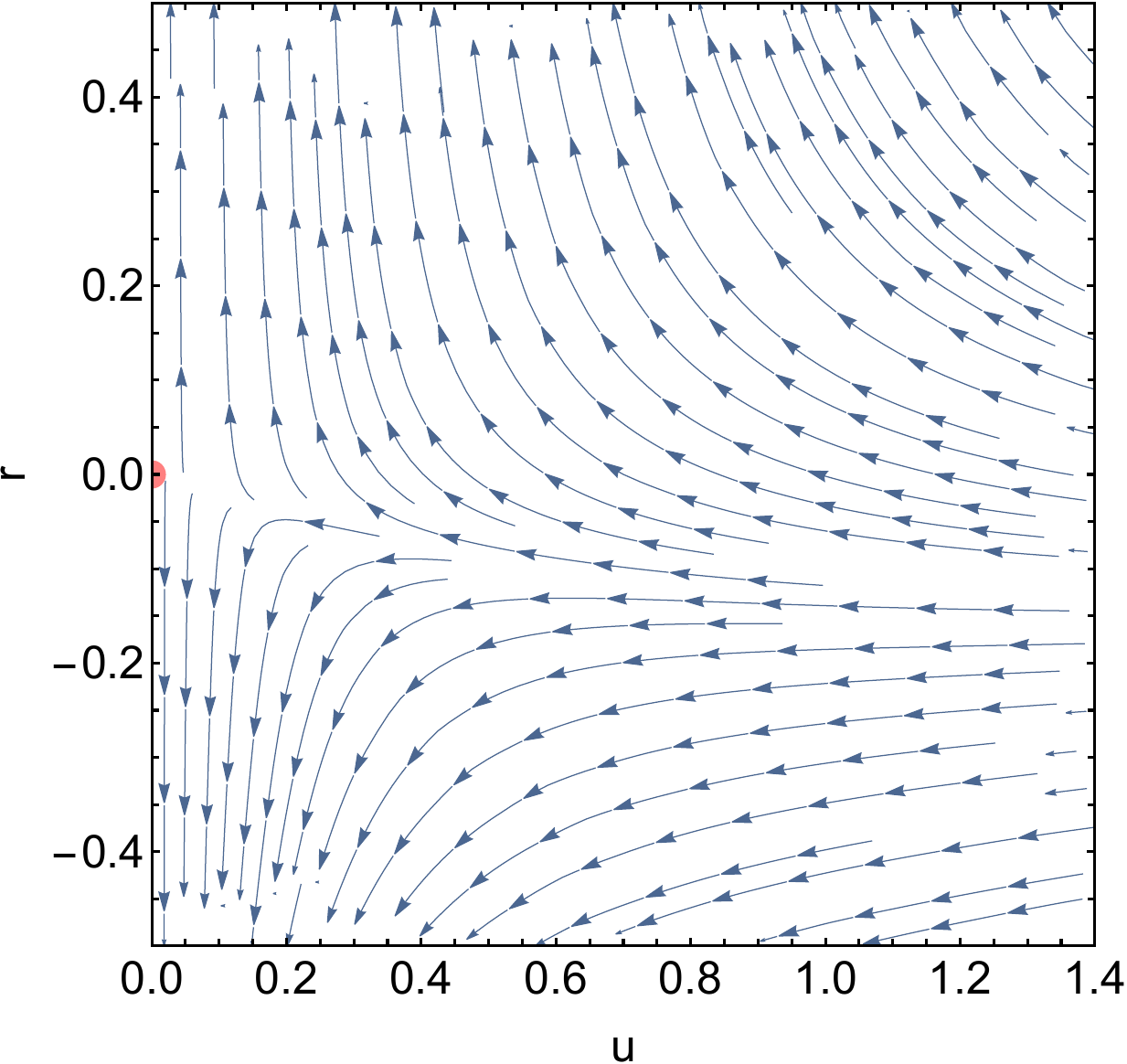}} 
\subfigure[]
		{\label{fig:FPgu4d}
			\includegraphics[width=0.31\textwidth]{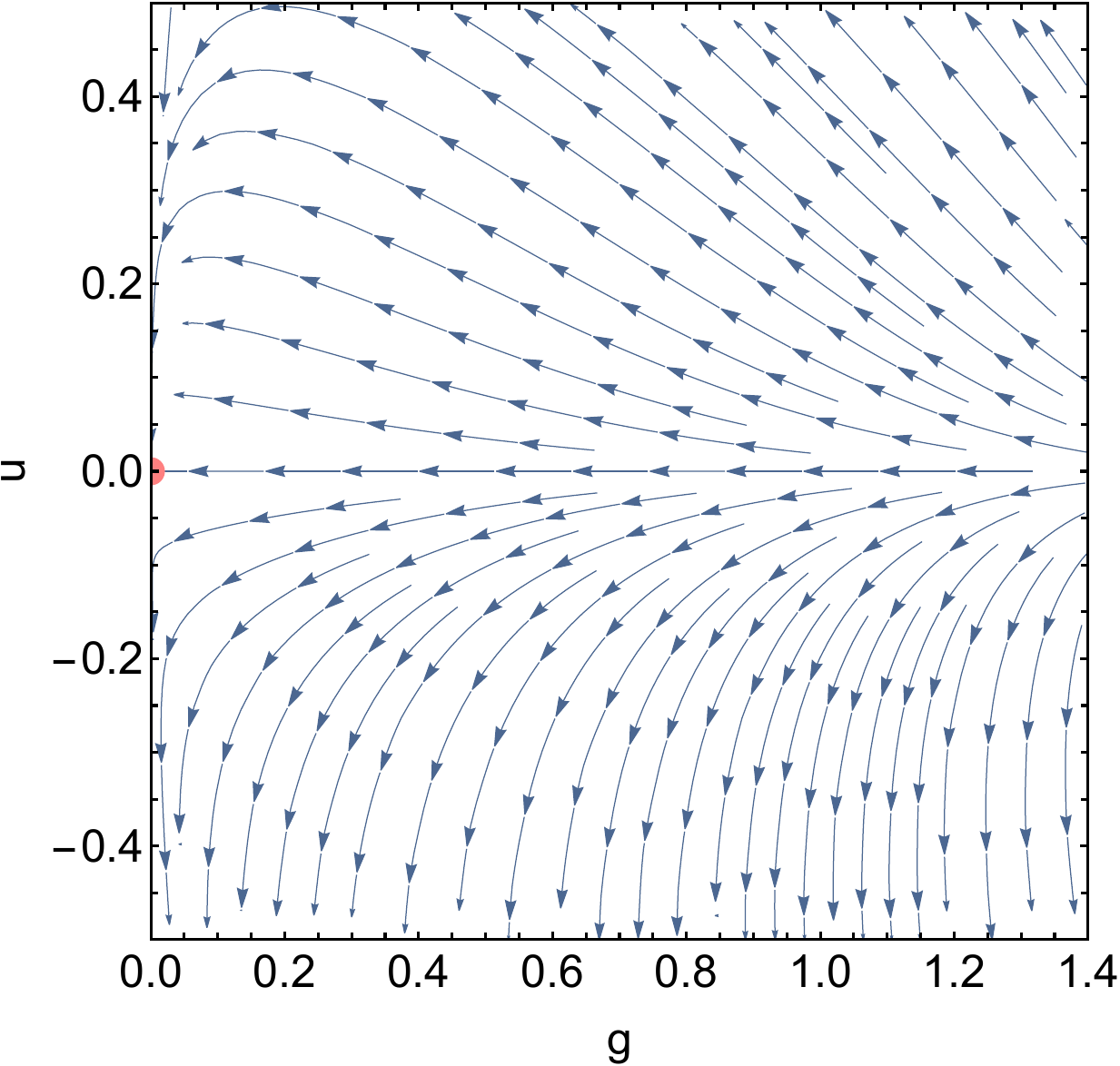}} 		
		\subfigure[]
		{ \label{fig:FPgr4d}
			\includegraphics[width=0.31\textwidth]{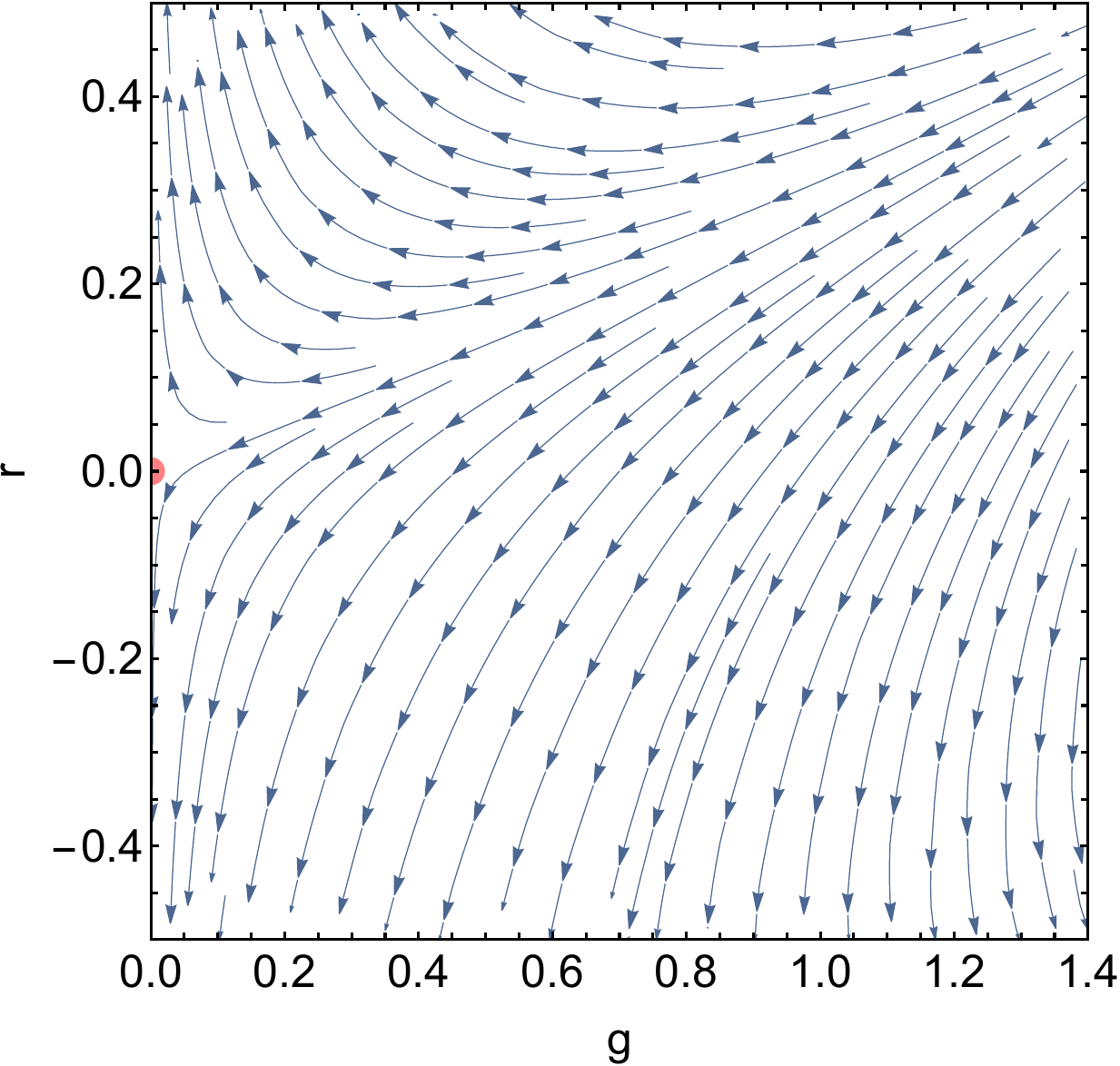}}
	\end{center}
	\caption{ DRG flux for $d\geq 4$, obtained by numerically solving Equations \ref{eq:r} to \ref{eq:g}.  We have fixed the Stratonovich prescription $\alpha=1/2$. The bold (red) dot  at $r=u=g=0$ is the Gaussian fixed point. We clearly observe that $r$ is a relevant scaling variable, while $u$ and $g$ are irrelevant.}
	\label{fig:flux4d}
\end{figure*}

For $d>4$,  $\epsilon<0$, thus there are  two negative eigenvalues $y^G_u<0,  y^G_g<0$ and a positive one, $y^G_r>0$.    The corresponding  eigenvectors  define a relevant scaling variable, $r$,   and a critical plane that divides the ordered and disordered phases,  whose normal vector is given by ${\bf n}_c={\bf v}_u^G\times {\bf v}_g^G$.  In Figure~\ref{fig:flux4d}, we show the flux diagram computed by numerically solving Eqs.~\ref{eq:r} to (\ref{eq:g}) for $d>4$.  
%%%%%%%%%%%%%%%%%%%
%%%%%%%%%%%%%%%%%%%%%%%%%%%%%%%%%%%%%%
We clearly see the stable Gaussian fixed point and the critical plane dividing both phases.  We observe that the numerical solution of the nonlinear equations correctly reproduces the behavior found in the linearization procedure (Eqs.~\ref{eq:yrG}-\ref{eq:vgG}) near the fixed point.   Summarizing, the critical properties of the model in $d>4$ are given by the usual Landau or mean field critical exponents  with an additive dynamics characterized by the dynamical critical exponent $z=2$. 

For $d<4$,   $\epsilon>0$
 and we have two positive eigenvalues $y^G_r>0, y^G_u>0$ and a negative one, $y^G_g<0$.   The eigenvectors are the same given by Eqs. (\ref{eq:vrgG})-(\ref{eq:vgG}),   but now there are two repulsive directions ($v_r^G$ and $v_u^G$) and just one direction attractive ($v_g^G$).   Therefore,   in this case, the fixed point is unstable and the flux runs away from it.  The phase transition is no longer ruled by the Gaussian fixed point and so, we need to look for another fixed point to describe the critical properties of the system.  Notice that there is an irrelevant direction with dimension $-2$  whose scaling variable is a line contained in the $rg$-plane ($u=0$),   with orientation  given by the vector of Eq. (\ref{eq:vgG}). Therefore,   any initial condition in this particular line in the $rg$-plane will be attracted by the Gaussian additive fixed point.  However, since the other two directions  $v_r^G$ and $v_u^G$ (given by Eqs.~\ref{eq:vrgG} and~\ref{eq:vuG}, respectively)  are repulsive,  any initial condition  away from this line (even infinitesimally near),  will produce a flux that depart from the Gaussian additive fixed point.

\subsubsection{The non-Gaussian Multiplicative  Fixed Point}

For $d<4$,  in addition to the unstable Gaussian additive fixed point, there is a nontrivial fixed point of Eqs.~\ref{eq:r}-(\ref{eq:g}), given at order $\epsilon^2$ by 
%%%%%%%%%%
\begin{eqnarray} 
r^*&=& -\frac{1}{6} \epsilon + \frac{(97+162\alpha)}{2916}\epsilon^2+ O(\epsilon^3)\; \label{eq:rstar} ,\\
u^*&=& \frac{2}{3} \epsilon - \frac{70}{729}\epsilon^2+ O(\epsilon^3)\; \label{eq:ustar}, \\
g^*& = & \frac{1}{18}\epsilon^2+ O(\epsilon^3) \label{eq:gstar}\; , 
\end{eqnarray}  
%%%%%%%%%%%
where $\epsilon=4-d$.    In the $ru$-plane, the position of the fixed point is similar to the Wilson-Fisher fixed point.   Although,  in our case, the position of the temperature coupling $r^*$ explicitly depends on the stochastic prescription $\alpha$.  Moreover,   we obtain a finite value of $g^*$ at order $\epsilon^2$.   That means that the effective dynamics at the critical point is no longer described by an additive Langevin equation but by a multiplicative one with dissipation function $G^2(\phi)=1+g^*\phi^2$. 

In order to explore the fixed point stability,   we linearize Eqs. (\ref{eq:r}) to (\ref{eq:g}) around this nontrivial fixed point. 
We define fluctuations of the coupling constants, 
\begin{eqnarray}
\delta r&= & r-r^* \; , \\
\delta u&=& u-u^* \; ,  \\
\delta g&=& g-g^* \; , 
\end{eqnarray}
 in such a way that the location of the fixed point is given by  $\delta r^*=\delta u^*=\delta g^*=0$.
 The linear DRG equations now read
 \begin{equation}
\left(
\begin{array}{c}
\frac{d\delta r}{d\ell} \\
\frac{d\delta u}{d\ell} \\
\frac{d\delta g}{d\ell} \\
\end{array}
\right)
=
\left(
\begin{array}{ccc}
T_{rr} & T_{ru} & T_{rg} \\
T_{ur} & T_{uu} & T_{ug} \\
T_{gr} & T_{gu} & T_{gg}
\end{array}
\right) 
 \left(
\begin{array}{c}
\delta r \\
\delta u \\
\delta g\\
\end{array}
\right)
\end{equation}
where the  matrix is given up to order  $\epsilon^2$  by
\begin{equation}
T=\left(
\begin{array}{ccc}
 2-\frac{1}{3}\epsilon +\frac{35}{729} \epsilon ^2 &\frac{1}{2}-\frac{5 }{36}\epsilon + \frac{(269-486 \alpha )}{17496} \epsilon ^2 & -2 \alpha  \\
  &  &  \\
 \frac{4 }{3}\epsilon ^2 & -\epsilon-\frac{62}{243} \epsilon ^2  & \frac{8 }{3}\epsilon-\frac{280}{729}\epsilon ^2 \\
&  &  \\
 0 & \frac{1}{3}\epsilon -\frac{35}{729} \epsilon ^2 & -2 \\
\end{array}
\right) +O\left(\epsilon^3\right).
\end{equation}

By diagonalizing $T$, we obtain the eigenvalues and eigenvectors that will define the relevant and irrelevant scaling variables. 
The eigenvalues are given by
\begin{eqnarray}
y_r&=&2 - \frac{\epsilon}{3} + \frac{278}{729} \epsilon^2 + O(\epsilon^3)
\label{eq:yr}\\
y_u&=&-\epsilon- \frac{35}{243} \epsilon^2+ O(\epsilon^3)
\label{eq:yu}  \\
y_g&=&-2 - \frac{4}{9}  \epsilon^2+ O(\epsilon^3)
\label{eq:yg}
\end{eqnarray}
from where it can be clearly seen that there is a repulsive direction related with the eigenvalue  $y_r>0$, and two attractive ones, $y_u<0$ and $y_g<0$.   

The directions of the scaling variables,  near the critical point,  are given by the corresponding normalized eigenvectors, 
\begin{eqnarray}
{\bf v}_r&=&\left\{1,0,0\right\}+\left\{0,\frac{2 }{3},0\right\}\epsilon ^2+O\left(\epsilon ^3\right) 
\label{eq:vr} \\
{\bf v}_u&=&\frac{1}{\sqrt{17}}
\left\{-1,4, 0 \right\}+ \frac{2}{3\sqrt{17}}
\left\{\frac{4 (12 \alpha +11) }{51},\frac{12 \alpha +11  }{51},1  \right\}\epsilon+
O\left(\epsilon ^2\right)
\label{eq:vu} \\
{\bf v}_g&=&\frac{1}{\sqrt{\alpha ^2+4}}\left\{\alpha,0,2\right\}+\frac{1}{{3 \sqrt{\alpha ^2+4}}}
\left\{\frac{\alpha +4  }{\alpha ^2+4},-8 ,-\frac{\alpha  (\alpha +4) }{2 \left(\alpha ^2+4\right)}\right\} \epsilon
+O\left(\epsilon ^2\right)
\label{eq:vg}
\end{eqnarray}
with $|{\bf v}_r|=|{\bf v}_u|=|{\bf v}_g|=1$ (to order $\epsilon^2$). While ${\bf v}_r$ indicates the relevant (repulsive) direction,   the critical plane defined by the normal vector ${\bf n}_c={\bf v}_u\times {\bf v}_g$ divides the parameter space in two regions  corresponding with the ordered and disordered phases.   

\begin{figure*}[htb]
	\begin{center}
		\subfigure[]
		{\label{fig:FPurS}
			\includegraphics[width=0.31\textwidth]{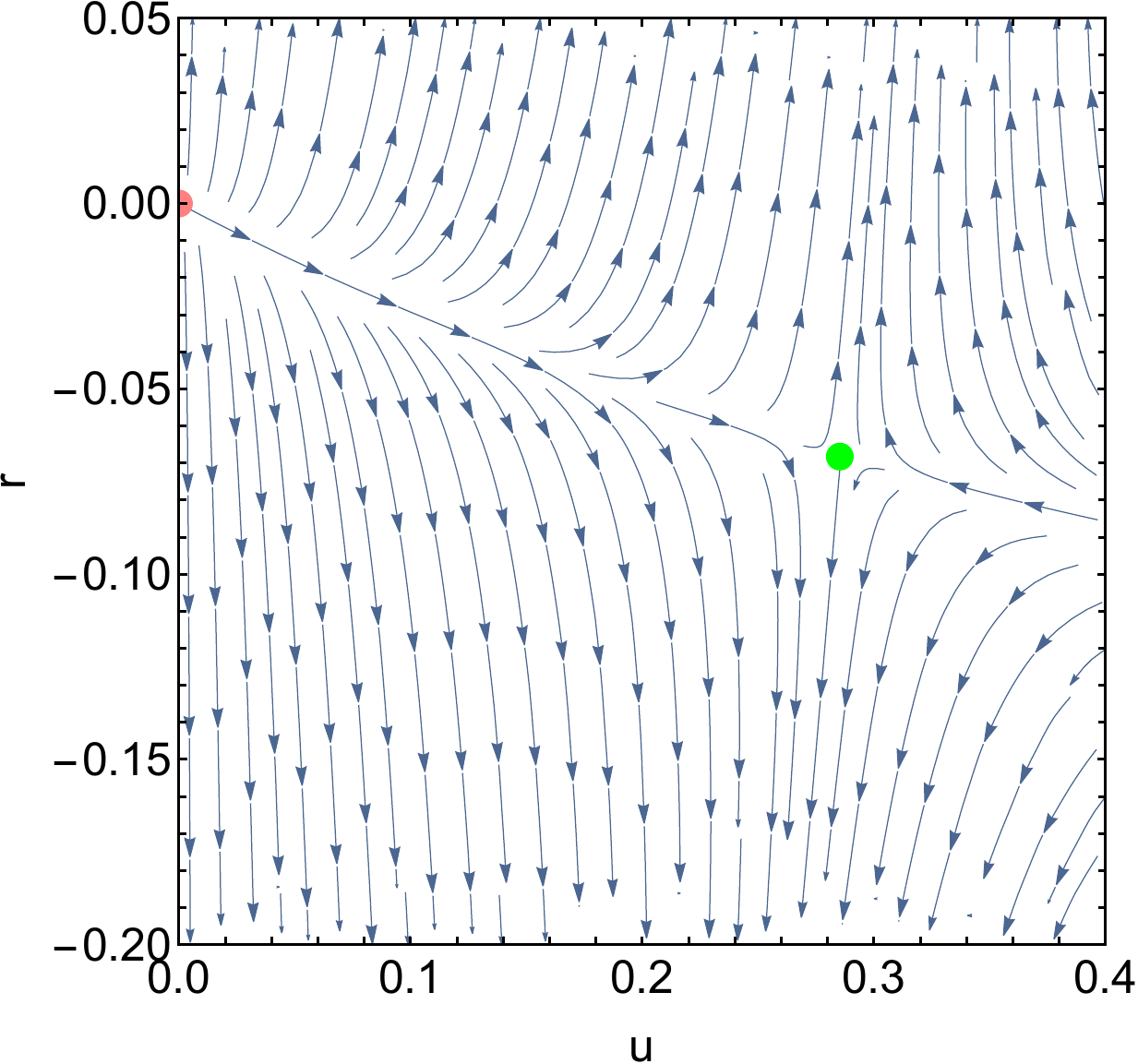}} 
		\subfigure[]
		{\label{fig:FPugS}
			\includegraphics[width=0.295\textwidth]{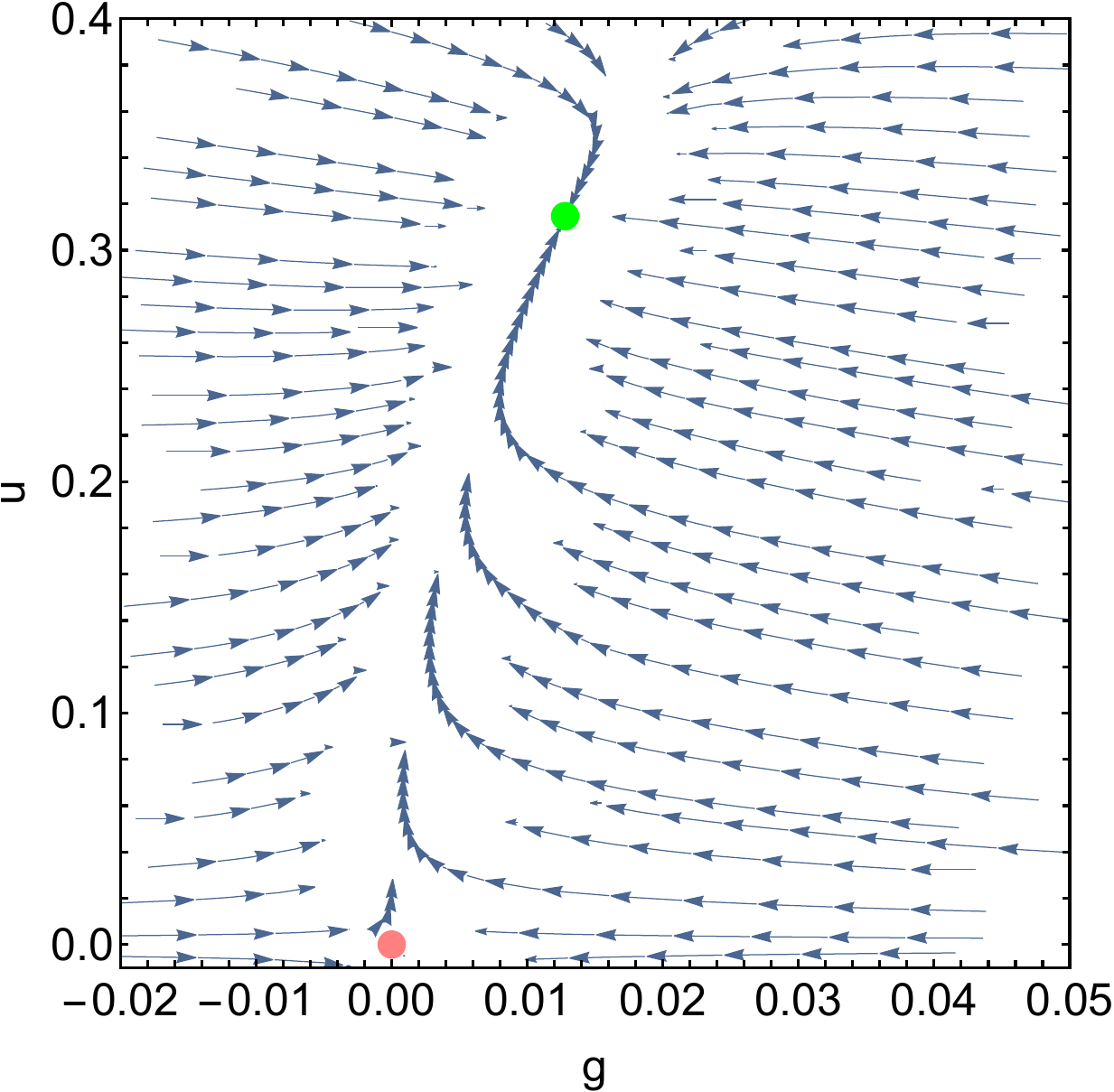}} 
		\subfigure[]
		{\label{fig:FPrgS}
			\includegraphics[width=0.31\textwidth]{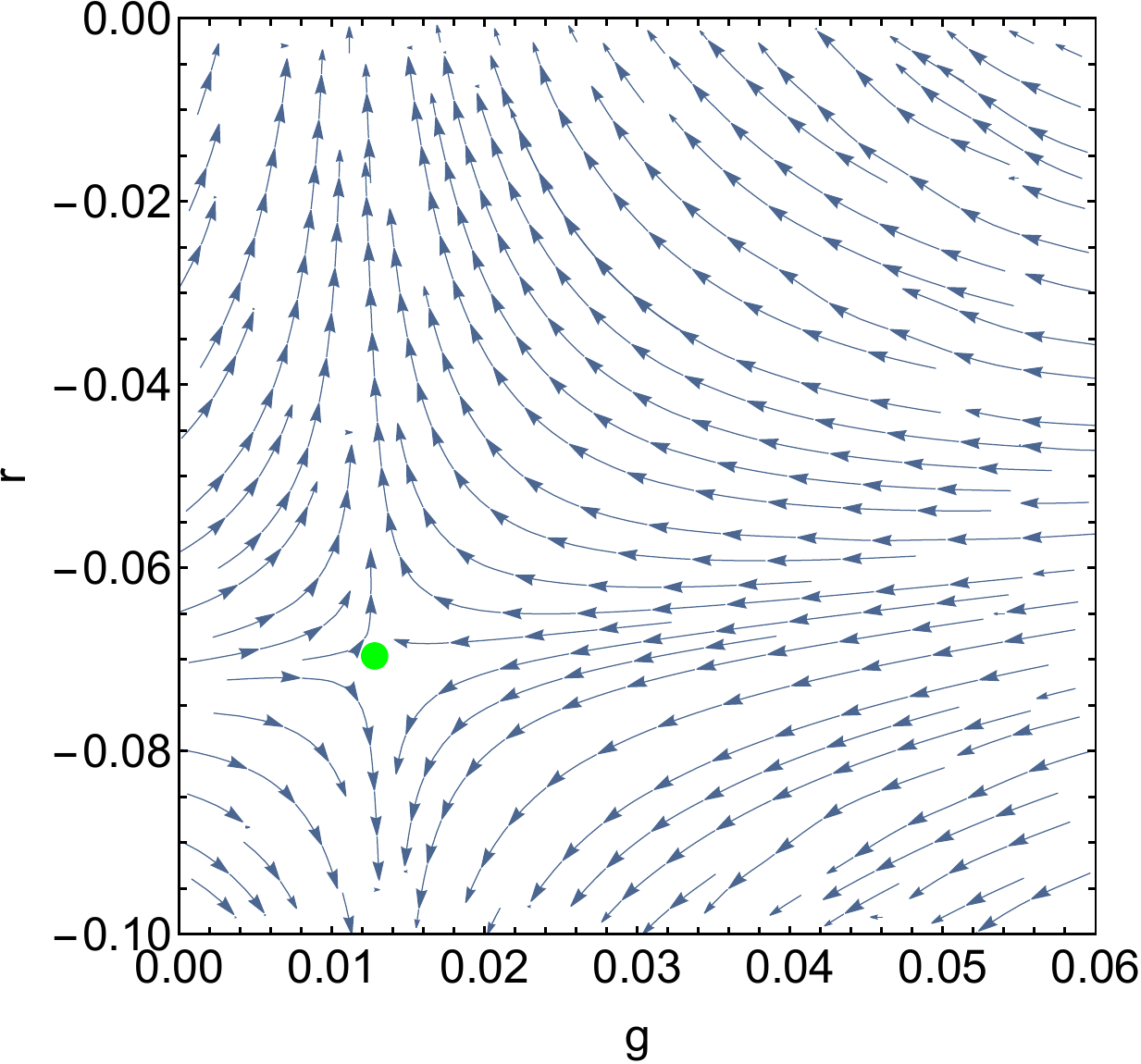}}
	\end{center}
	\caption{ DRG flux, obtained by numerically solving Equations \ref{eq:r} to \ref{eq:g} with $\epsilon = 0.5$.  We have fixed the Stratonovich prescription $\alpha=1/2$.  In (a),  we depict the flux in the $ru$-plane at $g^*=\frac{1}{18}\epsilon^2$.  It can be clearly seen a fixed point with one attractive and one repulsive direction with a similar topology of a Wilson-Fisher fixed point.   In (b), we show the $ug$-plane.  We can observe  a completely attractive fixed point in green and the Gaussian fixed point in pink with one attractive and one repulsive direction. In (c),  we show the fixed point  in the $rg$-plane.}
	\label{fig:DRGfluxesSTR}
\end{figure*}

We depict,  in  Figure~\ref{fig:DRGfluxesSTR},  the flux diagram obtained by numerically solving equations~\ref{eq:r}-(\ref{eq:g}).  In the figures, we have chosen  the Stratonovich prescription ($\alpha=1/2$) and we have fixed $\epsilon=0.5$.  We indicate with a bold (red) point the Gaussian fixed point and with a green one the non-Gaussian multiplicative fixed point.     
%%%%%%%%%%%%%%%%%%%
%%%%%%%%%%%%%%%%%%%%%%%%%%%%%%%%%%%%%%
We observe that the position of the fixed point as well as the topology of the flux completely agree with the previous description using the linear approximation.     From Figure~\ref{fig:FPugS},   it can be clearly observed that there is an  attractive line to the additive Gaussian fixed point,  provided $u=0$, in accordance with the linearization around the Gaussian fixed point previously described in \S\ref{subsub:Gaussian}.  However, no matter how small is $u$, the flux begins approximating the Gaussian fixed point up to a distance of order $u^2$ and afterwards, the flux deviates to the non-Gaussian fixed point.  At the stem of this behavior is the $u^2$ correction to the recurrence DRG equation (\ref{eq:g}).

We also checked the dependence with the stochastic prescription, obtaining a complete agreement between the linear approximation and the numerical solution near the critical point. 
One important observation is that, while the position of the fixed point depends on the stochastic prescription $\alpha$,  the eigenvalues and therefore the critical exponents  do not  depend on it.   Thus,  different dynamics produced by different stochastic prescriptions are in the same universality class. This is an important nontrivial result on multiplicative critical dynamics.

From equation (\ref{eq:gamma}),   we can read the dynamical critical exponent $z$, by demanding that $\Gamma$ be marginal.   Thus, 
\begin{equation} 
z = 2 +\frac{u^{*2} }{54} \; .
\label{eq:z}
\end{equation}
Using the value of the fixed point, Eq.  (\ref{eq:ustar}),   we find $z=2+(2/243) \epsilon^2+O(\epsilon^3)$.  It is interesting to note that, at this level of approximation, the multiplicative coupling constant does not affect the dynamical critical exponent.  In fact, the next correction in equation (\ref{eq:gamma}) is of order $g u^2$,  that contributes with an $O(\epsilon^3)$ term.  Details of this calculation can be found in Appendix~\ref{App:DRG-equations}.

On the other hand,   the multiplicative coupling indeed contributes for the anomalous dimensions.  We have found
\begin{eqnarray}
\eta_1&=  &  g^*+ \frac{2}{27}u^{*2}  \label{eq:eta1} \; ,  \\
\eta_2&= &   g^* -\frac{1}{27}u^{*2}\label{eq:eta2} \; .
\end{eqnarray}
From a technical point of view,  while the contribution of $g$ comes from  tadpole one-loop diagrams,   the term proportional to $u^2$ is a two-loop correction.   Notice that both terms are  of the same order $\epsilon^2$.  
An interesting observation is that,  since the first term of the r.h.s of both Eqs. (\ref{eq:eta1}) and (\ref{eq:eta2}) are equal, 
$g^*$ does not affect the difference $\eta_1-\eta_2$,  which is  proportional to the static anomalous dimension $\eta$ computed in Ref.~\cite{Hohenberg-Halperin-1977}.
In addition,  we will show in the next section that $g^*$ does not correct  the dynamical critical exponent $z$ either  
(please see  the next subsection \S\ref{subsection:scaling} and  Appendix~\ref{App:Rescaling} for details).
These results are of order $\epsilon^2$. The multiplicative coupling $g$ will surely contribute to $z$ at order $\epsilon^3$.      

There is a last point that we would like to address to complete the calculation of the $\epsilon$ expansion. 
Equations from (\ref{eq:r}) to (\ref{eq:gamma}) were computed by considering only the first multiplicative coupling $g\equiv g_1$ in equation  (\ref{eq:GdeltaG}).  Thus, it is important to check how the higher order couplings $g_n$, with $n>1$ modify our result. Higher order terms produce vertices with two dashed legs, corresponding to the response variable $\varphi$, and an increasing number of legs corresponding with the power of the order parameter $\phi$. This fact strongly constrains the number of loop diagrams that contribute to the DRG equations.  The result is that $g_n$ with $n\ge 2$ are completely irrelevant at the nontrivial fixed point of Eqs. (\ref{eq:rstar})-(\ref{eq:gstar}),   at least at order $\epsilon^2$.  It could happen that, for higher orders in the $\epsilon$ expansion,  the fixed point could acquire nontrivial values of $g_2,  g_3$ etc.    However, if this is going to occur, it will be at order greater than $\epsilon^3$. 
In Appendix~\ref{App:g2}, we give explicit details of this calculation. 

%%%%%%%%%%%%%%
\subsection{Corrections to  scaling in multiplicative systems}
\label{subsection:scaling}
%%%%%%%%%%%%%%
The scaling properties of the fields in momentum and frequency,  displayed in Eqs.~\ref{eq:varphiscaled}) and (\ref{eq:phiscaled}),  can be rewritten in configuration space by just Fourier transforming Eq.  (\ref{eq:FourierPhiM}).  The order parameter and the response field scale as
\begin{eqnarray}
\phi^<(b{\bf x},b^zt) &= & b^{-\frac{d-2-\eta_2}{2}}\phi({\bf x},t) \; ,  \\
\varphi^<(b{\bf x},b^zt)& = &  b^{-\frac{d+2+\eta_1}{2}}\varphi({\bf x},t)  \; .
\end{eqnarray}
$z$ is the dynamical critical exponent and $\eta_1$ and $\eta_2$ are the anomalous dimension of the order parameter and the response field, respectively (the sign difference between $\eta_1$ and $\eta_2$ is just a matter of convention).  

The scaling hypothesis tells us that,  at very long times and long distances, the dynamical correlation function should scale as
\begin{equation} 
	G(b({\bf x}-{\bf x}'),b^z(t-t')) =\langle\phi(b {\bf x},b^z t)\phi(b {\bf x}',b^z t')\rangle= b^{-(d-2-\eta_2)}G({\bf x}-{\bf x}',t-t') \; .
	 \label{eq:G_scaling}
\end{equation}
On the other hand, the response function
\begin{equation} 
	R(b({\bf x}-{\bf x}'),b^z(t-t')) =\langle\varphi(b {\bf x},b^z t)\phi(b {\bf x}',b^z t')\rangle= b^{-\left(d +\frac{\eta_1-\eta_2}{2}\right)}R({\bf x}-{\bf x}',t-t') \; .
	 \label{eq:R_scaling}
\end{equation}
Therefore, in principle, the scaling hypothesis is given in terms of three independent parameters $\{z, \eta_1,\eta_2\}$. 
However, if we assume that the system converges to equilibrium at long times, the linear response and the correlation functions are related by the Fluctuation--Dissipation Theorem  (Eq.  (\ref{eq:FDT2})).  Therefore, $\{z, \eta_1,\eta_2\}$ are no longer independent. 

For additive stochastic processes,  the susceptibility  $\chi({\bf x}-{\bf x}',t-t')\sim R({\bf x}-{\bf x}',t-t')$ (See Eq. (\ref{eq:response-Add})).  Then, replacing  Eqs.~\ref{eq:G_scaling} and~\ref{eq:R_scaling} into Eq. (\ref{eq:FDT2}),  we immediately find, 
\begin{equation}
z=2+\frac{\eta_1+\eta_2}{2}  \;. 
\end{equation}   

However, for multiplicative stochastic processes, the susceptibility is no longer proportional to $R({\bf x}-{\bf x}',t-t')$ but has corrections coming from the diffusion function $G(\phi)$, given by Eq. (\ref{eq:response}).   Then,  the FDT is given by a more involved relation
\begin{eqnarray}
i \frac{\Gamma}{2}\left\langle \phi({\bf x}, t) G^2(\phi({\bf x}'))\varphi({\bf x}',t') \right\rangle -\frac{\Gamma}{2} \left\langle \phi({\bf x},t)\frac{\delta G^2(\phi)}{\delta\phi({\bf x}',t')}\xi({\bf x}',t')\bar\xi({\bf x}',t') \right\rangle
= - (\partial_t - \partial_{t'})\langle\phi( {\bf x},t)\phi({\bf x}', t')\rangle \Theta(t-t').
  \label{eq:FDT3}
\end{eqnarray}
Computing the correlators perturbatively in $g$  and using the scaling of Eqs. (\ref{eq:G_scaling}) and (\ref{eq:R_scaling}), we obtain 
\begin{equation}
z=2+\frac{\eta_1+\eta_2}{2}-g^*  \;. 
\label{eq:z-corrected}
\end{equation}   
We can observe that multiplicative noise changes the scaling properties of the linear susceptibility,  modifying the usual constraint between the dynamical critical exponent and the anomalous dimensions. 

By replacing Eqs. (\ref{eq:eta1}) and (\ref{eq:eta2})  into Eq.  (\ref{eq:z-corrected}),   we find 
\begin{equation} 
z = 2 +\frac{u^{*2} }{54}
\end{equation}
which is the result (independent of $g^*$) obtained  in  Eq. (\ref{eq:z}),  computed by direct methods.  

Thus,   we conclude that  the perturbative calculations of $z$, $\eta_1$ and $\eta_2$  computed in Eqs. (\ref{eq:z}), 
(\ref{eq:eta1}) and (\ref{eq:eta2}) are consistent with the dynamical scaling hypothesis and the equilibrium properties codified in the FDT. 

It is worth  to mention that  the tadpole diagrams arising from the first term of Eq. (\ref{eq:FDT3}) exactly cancel the tadpole diagrams coming from the Grassmann fields in the second term.   This is important because these diagrams carry the information about the stochastic prescription $\alpha$ and,  as we have already mentioned,   $z$, or any other critical exponent,  should not depend on $\alpha$ since different prescriptions are in the same universality class.  

%%%%%%%%%%%%%%%%%%%%%%%%%
\section{Summary and discussions}
\label{Sec:Discussions}

We have shown different aspects  of multiplicative noise stochastic processes,  specially those relevant to describe dynamics of phase transitions. 

In Section~\ref{Sec:Multiplicative-Noise},  we have presented a brief review of multiplicative noise stochastic processes.  We have specially focused on equilibrium properties.   In particular, we have discussed the   conditions to reach an  equilibrium state at long times. In general,  ``stochastic equilibrium" is defined by zero probability current,  $J(\phi)=0$ (where $J(\phi)$ is given by Eq. (\ref{eq:Jex}) ).  This condition leads to  an equilibrium probability distribution,  given by Eq. (\ref{eq:Peqdistr})   that,  in general,  is not of the Boltzmann type.  There are two cases where this distribution reduces to the Boltzmann one:  for usual  additive noise ($G(\phi)=1$),   and for multiplicative noise in the H\"anggi-Kimontovich or thermal prescription, $\alpha=1$.  Therefore,   ``thermodynamic equilibrium" is a special case of ``stochastic equilibrium"\cite{Arenas2012-2}.

For weak noise,  or  for perturbative multiplicative corrections to additive noise,  we expect that temperature and coupling constants be smoothly corrected,  without affecting  the topology of the phase diagram.  
 However, as we have shown in Eqs. from (\ref{eq:UeqCorrected}) to (\ref{eq:vtilde}),   the multiplicative dynamics could completely change the static properties of the system, even changing the order of the phase transition. It is interesting to note that this dramatic change can happen for any choice of the stochastic prescription except for the thermal prescription $\alpha=1$.   Thus, in some sense,  thermal equilibrium protects the order of the phase transition.  It appears that when the state is significantly distant from thermal equilibrium,  the impact of multiplicative noise on altering the  order of the phase  transition becomes more pronounced; been maximum at the It\^o prescription ($\alpha=0$).  In this way,  we do not expect this effect to take place in thermodynamic systems,  where equilibrium is governed by the Boltzmann distribution. 

When dealing with symmetries and fluctuations theorems,  it is important to have a formalism that treats all stochastic prescriptions in the same footing. This is because,  time reversal transformation mixes different prescriptions. 
For this reason,  we have generalized the formalism developed in Refs.  \cite{Arenas2012-2, Miguel2015,arenas2010,Arenas2012} to treat continuous infinite degrees of freedom, appropriated to study dynamics of phase transitions. 
We have presented, in Section~\ref{Sec:Formalism}, the functional formalism  used to compute dynamical correlation and response functions.   We showed how to compute linear response and the specific form of the {\em Fluctuation--Dissipation Theorem} for multiplicative stochastic processes.  Since the FDT is an exact result, it is very useful to check any approximation scheme,  in particular the $\epsilon$-expansion.  

The core of the paper is contained in  Section~\ref{Sec:DRG}, where we have built the {\em Dynamical Renormalization Group} equations in the $\epsilon$ expansion approximation. We have analyzed a model of a real order parameter driven by a multiplicative Langevin equation.  In particular, we have considered a model with $Z_2$ symmetry, where $H(\phi)=H(-\phi)$ and $G(\phi)=G(-\phi)$.  Equations  from (\ref{eq:r}) to (\ref{eq:gamma}) are the main result of the paper. They determine the flux diagram in the  parameter space $\{r,u,g\}$.    We have performed a perturbative calculation up to second order in the couplings, and we have retain only terms that contribute to order $\epsilon^2$. 
 
We have numerically solved these equations and  we  have presented the results for $d>4$ in Figure~\ref{fig:flux4d}. 
We observed the stable Gaussian fixed point  $r^*=u^*=g^*=0$. It is clear that  $r$ is a relevant scaling variable,   while $u$ and $g$ are irrelevant, defining the critical plane that separates the ordered and disordered phases. From the dynamical point of view,  the system is well described by an {\em additive Langevin equation}.  

For $d<4$,  the Gaussian fixed point ($r^*=u^*=g^*=0$) becomes unstable, and the flux goes to a novel fixed point with $u^*\neq 0$ and $g^*\neq 0$.   We show the flux diagram for this case in Figure~\ref{fig:DRGfluxesSTR}.     We have computed the fixed point to order $\epsilon^2$ in Eqs. (\ref{eq:rstar}), (\ref{eq:ustar}) and (\ref{eq:gstar}).   We observed that the phase transition is dominated by  this novel fixed point, driven by a multiplicative noise dynamics.   At this order of approximation,  the dynamics very near the phase transition is driven by a {\em multiplicative Langevin equation} with diffusion function,   $G(\phi)=1+ g^* \phi^2/2$, with  $g^*\sim \epsilon^2$. This is our main result.

This unexpected fact opens several interesting fundamental questions.  On the one hand, the static critical exponents are very well known and have been computed by several method since the pioneer work of Wilson and Fisher\cite{Wilson-1972}.   Of course,  our results,  obtained with dynamical methods,  coincide with them.  Multiplicative couplings can modify non-universal features of the transition, such as the critical temperature, but they do not  modify the universal behavior.  The dynamical critical exponent is more involved. Since it is not protected by symmetries and dimensionality, is not clear what will happen at higher orders. Our computation of $z$ is compatible with the existing body of numerical simulations.  Moreover, the subtle cancellations that makes $z$ independent of $g^*$ are proper to the order of the approximation.  We cannot guarantee these cancellations at higher orders; indeed,  it is probable that they do not occur.  Moreover, for orders greater than  $\epsilon^2$, higher multiplicative couplings will appear, and the fate of the dynamical critical exponent is hard to predict. Additionally, the possibility of encountering crossovers at various time scales cannot be discounted. It's worth noting that these effects fall beyond the intended scope of the current paper.

There is an important question that arise from the fact that multiplicative noise is been generated by DRG transformations.  Provided we are describing a physical system in thermodynamic equilibrium,   we are force to fix the thermal prescription $\alpha=1$,   since this prescription leads to the Boltzmann distribution.  
However,  we found that  critical exponents do not depend on the prescription (at least at order $\epsilon^2$);  different dynamics,  driven by different stochastic prescriptions (for instance,  It\^o,  Stratonovich,  H\"anggi-Klimontovich, etc.),  are in the same universality class.  For this reason, the description of equilibrium dynamics by using Langevin equations makes sense.   

It is very difficult to speculate which modifications our results will suffer at higher orders in the 
$\epsilon$ expansion.  We have checked that at order $\epsilon^2$,  higher order terms in the diffusion function are irrelevant.   However,  for greater orders,  it is possible that the multiplicative fixed point could flow to  a diffusion  function,   $G^*(\phi)$,   with higher powers of $\phi$.   
A very interesting question is which is the fate of the diffusion function at the critical point at lower dimensions.  Interestingly,   in the extreme case $d=2$,  the whole diffusion function $G(\phi)$ is marginal at tree level.  However,  at this dimension, the $\epsilon$-expansion is not reliable.    We leave the study of these questions for a future presentation.

\section*{Acknowledgments}
We would like to acknowledge  Zochil Gonz\'alez Arenas for a careful reading of the manuscript and useful comments. 
The Brazilian agencies, {\em Funda\c c\~ao de Amparo \`a Pesquisa do Estado do Rio
de Janeiro} (FAPERJ), {\em Conselho Nacional de Desenvolvimento Cient\'\i
fico e Tecnol\'ogico} (CNPq) and {\em Coordena\c c\~ao  de Aperfei\c coamento de Pessoal de N\'\i vel Superior}  (CAPES) - Finance Code 001,  are acknowledged  for partial financial support.  NS was partially supported by a ``Sandwich  PhD Fellowship"  from the program CAPES-PrInt at CASUS.

%%%%%%%%%%%%%%%%%%%%%%%%%%%%%%%%%%%
\appendix
%%%%%%%%%%%%%%%%%%%%%%%%%%%%%%%%%%%

%%%%%%%%%%%%%%%%%%%%%%%%%%%%%%%%
\section{DRG-equations}
\label{App:DRG-equations}
%%%%%%%%%%%%%%%
The first step to build the perturbative DRG equations  is to compute $S'_{\Lambda/b}$ using Eq.   (\ref{eq:Sperturbative}).  Then, we rescale momentum, frequency and fields using Eqs. (\ref{eq:bk})-(\ref{eq:xiscaled}), 
and finally we compare $S'_\Lambda$ with $S_\Lambda$.   
All the terms in the computation of Eq. (\ref{eq:Sperturbative}) can be cast in terms of Feynman diagrams sketched in  Figures \ref{fig:1V1loop}, \ref{fig:2V2loop}, \ref{fig:u2_1loop}, \ref{fig:ug_1loop} and  \ref{fig:g2_1loop}.  In Appendix~\ref{App:Integrals}, we explicitly show how to compute all the loop integrals and in Appendix \ref{App:Rescaling},  we  explicitly make the rescalings.  We have computed all diagrams at second order in the couplings $u$ and $g$  and, after rescaling,  we have found the following DRG equations:
\begin{eqnarray} 
\frac{dr}{d\ell} &=&2r+\frac{u}{2 (r+1)}-2 \alpha  g+\frac{2 g u}{3 (r+1)^2}-\frac{u^2}{6 (r+1)^3}+\frac{ru (3 d (r+1)-4) (4 g (r+1)-u)}{54 d (r+1)^5}  \label{app:r} 
\\
	\frac{du}{d\ell}&\!\!=&\!\!\!\!\!\!u \left(4-d-\frac{3 u}{2 (r+1)^2} +\frac{4 g}{r+1}-\frac{2 g u}{3 (r+1)^3}+\frac{u^2}{18 (r+1)^4}+\frac{ u(9 d (r+1)-8) (4 g (r+1)-u)}{54 d (r+1)^5}\right)
	\label{app:u} \\
	\frac{dg}{d\ell}& =&g \left(2-d +\frac{g}{r+1}-\frac{5 u}{2 (r+1)^2} +\frac{u (3 d (r+1)-2) (4 g (r+1)-u)}{27 d (r+1)^5}\right) +\frac{u^2}{4 (r+1)^3}
	\label{app:g}\\
\frac{d\Gamma}{d\ell}&=& \Gamma  \left(z-2-\frac{2 u^2}{27 d (r+1)^5}+\frac{8u g}{27 d (r+1)^4} \right)\label{app:gamma}
\end{eqnarray}
%%%%%%%%%%%%%%
In these equations, all couplings  $r$, $u$ and $g$ are dimensionless since we have rescaled them as 
\begin{eqnarray}
\Lambda^{-2} \; r&\to& r  \; ,  \\
\frac{\Omega_d}{(2\pi)^d} \; \Lambda^{d-4} \;  u&\to & u \; ,    \\
\frac{\Omega_d}{(2\pi)^d}\; \Lambda^{d-2}  \; g&\to &  g \; , 
\end{eqnarray} 
where $\Omega_d$  is the area of a $(d-1)$-dimensional sphere of radius one. 

This expansion is meaningful,  in the sense that it is controlled,  provided we work near the upper critical dimension $d_c=4$.  Then, we define the small variable $\epsilon=4-d$ and look  for fixed points making an expansion to order $\epsilon^2$. 
For this, we look for solutions of  $dr/d\ell=du/d\ell=dg/d\ell=d\Gamma/d\ell=0$,   making an expansion of the form
%%%%%%%%%%%%%%%%%%%
\begin{eqnarray}
    r^{*} &=& r_1^* \epsilon +r_2^*\epsilon^2 + O(\epsilon^3)
    \\
    u^{*} &= &  u_1^* \epsilon +u_2^*\epsilon^2 + O(\epsilon^3)
    \\
    g^{*} &= &  g_1^* \epsilon +g_2^*\epsilon^2 + O(\epsilon^3)
\end{eqnarray}
%%%%%%%%%%%%%%%%%%%
By comparing the coefficients with equal order of $\epsilon$, we find, 
\begin{eqnarray}
r^{*}_1&=&-\frac{1}{6}\mbox{~~~,~~~} r^{*}_2=\frac{97+162\alpha}{2196}  \\
u^{*}_1&=&\frac{2}{3}\mbox{~~~~~,~~~} u^{*}_2=-\frac{70}{729}  \\
g^{*}_1&=&0\mbox{~~~~~~,~~~} g^{*}_2=-\frac{1}{18} 
\end{eqnarray}

Since $g_1^*=0$,  the multiplicative dynamical fixed point appears at order $\epsilon^2$.    Therefore,   by expanding the terms  $1/(1+r)$ in powers of $r$ and keeping only the terms that contribute to order $\epsilon^2$,   we find the  simplified system given by Eqs. (\ref{eq:r})-(\ref{eq:gamma}).

%%%%
\section{Effect of higher order multiplicative couplings}
\label{App:g2}

To define the multiplicative Langevin dynamics, we have considered a general diffusion function of the form
\begin{equation}
G^2(\phi)=1+ g \phi^2+g_2 \phi^4+ g_3\phi^6+\ldots
\label{app:G2}
\end{equation}
To build the DRG equations we have only considered the first coupling constant $g$ and we have ignored 
the higher order couplings $g_2, g_3, \ldots$.   The argument was that higher order couplings are irrelevant and can be ignored.   This is true at order $\epsilon^2$. 

\begin{figure}
	\begin{center}
\includegraphics[width=0.5\textwidth]{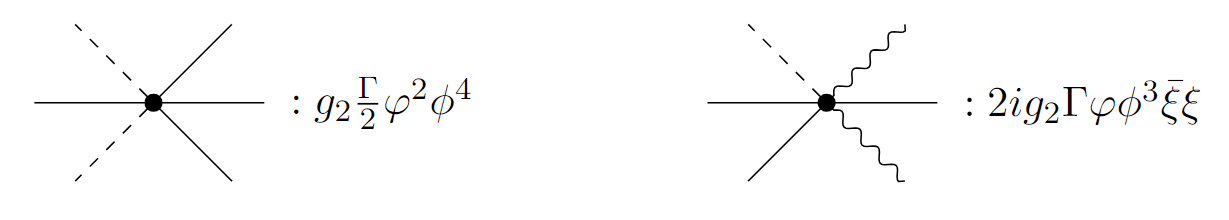}
	\end{center}
	\caption{Vertices induced by the coupling constant $g_2$.}
	\label{fig:vertexG2}
\end{figure}

To see this,   let us consider, in addition to $g$, the higher order coupling $g_2$. This coupling introduces two additional vertices to the perturbative expansion, as shown in Figure~\ref{fig:vertexG2}.
%%%%%%%%%%%%%%%%%

\begin{figure}
	\begin{center}
\includegraphics[width=0.55\textwidth]{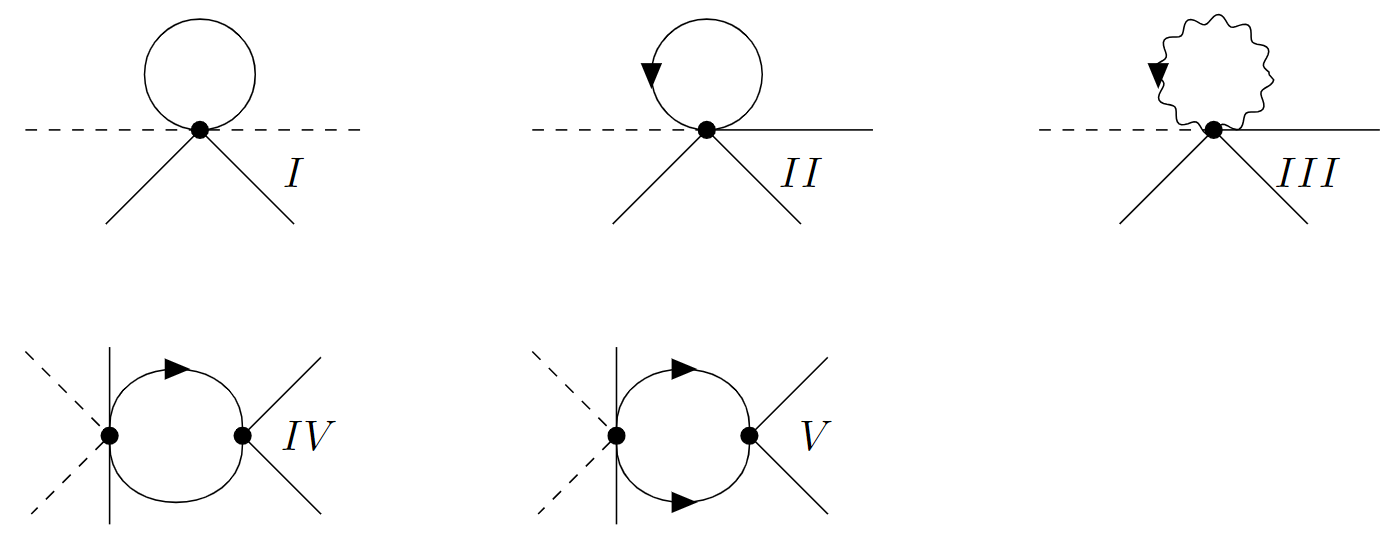}
	\end{center}
	\caption{One loop contributions of the new Bosonic vertex proportional to $g_2$.}
	\label{fig:loopG2}
\end{figure}

The new 1-loop contributions to the Bosonic sector are depicted in Figure~\ref{fig:loopG2}.

%%%%%%%%%%%%%%%%%%

\begin{figure}[htb]
\begin{center}
\includegraphics[width=0.45\textwidth]{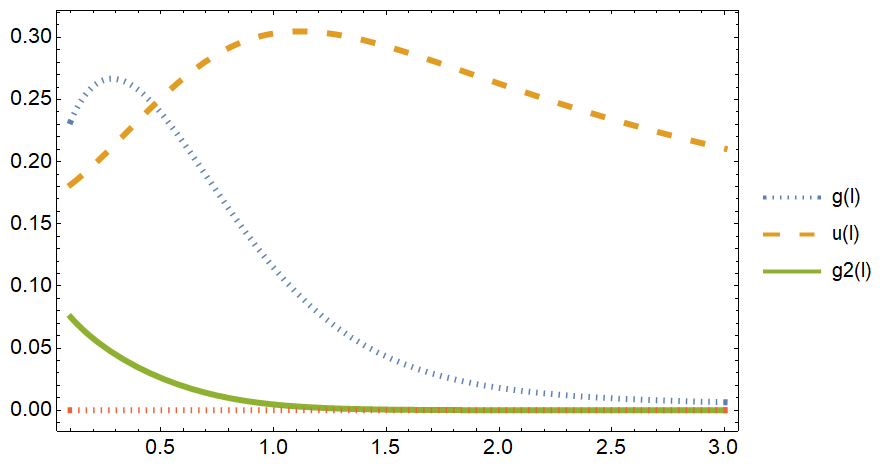}
\end{center}
\caption{Numerical solution of the DRG equations, Eqs. (\ref{app:DRG-r}) to (\ref{app:DRG-g2}) in $d=4-\epsilon$ and $\alpha=0$ for initial conditions $u(0)=g_1(0)=0.17$ and $g_2(0)=0.1$. 
The dashed line represents the flow of $u(\ell)$, the dotted line, $g_1(\ell)$, while $g_2(\ell)$ is represented by the solid line.}
\label{fig:flowBellow}
\end{figure}

One immediate consequence of these diagrams is that they do not renormalize $r$. Because of the relation between the Grassmann and the response field propagator codified in Eqs. \ref{eq:Ralpha} and \ref{eq:grassmannpropagator}, diagrams $II$ and $III$ cancel each other, meaning that there are also no corrections to coupling constant $u$. Therefore, $g_2$ only contributes to modify  $dg/d\ell$. Thus, the set of DRG equations at order $\epsilon^2$ are:
\begin{eqnarray}
	\frac{dr}{d\ell}& =&2r+\frac{u(1-r)}{2}-\frac{u^2}{6}-2 \alpha  g  
	\label{app:DRG-r}
	\\
\frac{du}{d\ell}&=&u\left[\epsilon-\frac{3 u(1-2 r)}{2}-\frac{2u^2}{27}+4 g \right]
	\label{app:DRG-u}
	\\
\frac{dg}{d\ell} &=&-2 g  +\frac{u^2}{4} +12g_2
	\label{app:DRG-g1}
\\
\frac{dg_2}{dl}& =&-4 g_2   
\label{app:DRG-g2}
	\\
\frac{d\Gamma}{d\ell}&=& \Gamma  \left(z-2-\frac{ u^2}{54}\right)
	\label{app:DRG-Gamma}
\end{eqnarray}
The only  possible solution for the fixed point is $g_2^* =0$,   even for the non Gaussian fixed point $u^*\neq 0$ and $g^*\neq 0$.   Moreover, from Eq. (\ref{app:DRG-g2}), it is trivial to realize that $g_2$ is irrelevant, with dimension $-4$.  In Figure~\ref{fig:flowBellow},  we show the numerical solution of the DRG equations near $d_c=4$ for initial conditions very near the Gaussian fixed point. 
%%%%%%%%%%%%%%%%%%%%%%
%%%%%%%%%%%%%%%%%%%%
As expected, we observe that the couplings $u, g$ evolve towards the nontrivial fixed point while   $g_2$ exponentially decays to zero.  For 
large $\ell$,  coupling constants  converge to the multiplicative noise fixed point characterized by $r^*,  u^*, g_1^*$ given by Eqs.  (\ref{eq:rstar})-(\ref{eq:gstar}),  and $g^*_2=0$. Therefore, the multiplicative noise fixed point,  computed in the main text of the paper,  is stable under $g_2$ perturbations.  In other words, $g_2$ is {\em irrelevant} with respect to the new fixed point.  In the figure, we can see that $g_2$ is a sub-leading variable and approaches zero very quickly, even before the other variables have reached the fixed point. 
Higher order multiplicative noise couplings produce vertices with an increasing number of legs. This fact strongly constrains the number of diagrams that can be constructed from these vertices and, at this order of approximation,  they will be more irrelevant than $g_2$.  Of course, in a higher order expansion in $\epsilon$,   equation (\ref{app:DRG-g2}) could be modified by higher order diagrams,  opening the possibility of a nontrivial fixed point for $g_2$. 

Summarizing, we can conclude that the diffusion function at the fixed point,  at order $\epsilon^2$,  is given by 
\begin{equation}
G^2(\phi)=1+g^* \phi^2.
\end{equation}
Higher order terms can only occur at higher orders in the $\epsilon$ expansion.

%%%%%%%%%%%%%%%
\section{Diagrammatic computations}
\label{App:Integrals}
%%%%%%%%%%%%%%%%%
In order to build the DRG equations, we found the effective action $S'(\phi^<, \varphi^<, \bar\xi^<, \xi^<)$ by computing all the one and two vertices diagrams depicted in Figures~\ref{fig:1V1loop}-\ref{fig:g2_1loop}. 

The single vertex diagrams in Figure~\ref{fig:1V1loop}  are essentially constants, since the tadpoles in configuration space mean correlations computed at a single time and position,  {\em i.e.},   $G^>(0, 0)$ or $R^>(0,0)$. 
In the usual treatment of quantum field theory, correlations at the same point and time are not well defined. In fact, the results often depend on the order of the limits $t\to 0$ and ${\bf x}\to 0$. However, in the Wilson approach to DRG,  the integrals over momentum are always cut-off in the ultraviolet as well as the infrared regimes. This fact makes this local quantities perfectly well defined.  Indeed, all the momentum integrals will be done in the infinitesimal spherical shell   $(1-d \ell) <|{\bf k}|/\Lambda<1$.
For instance,  using Eq.  (\ref{eq:R+}),  the bare response function in configuration space is 
\begin{eqnarray}
R^>({\bf x}, t)&=&\int_{1-\delta \ell <|{\bf k}|/\Lambda<1} \frac{d^dk}{(2\pi)^d}\int_{-\infty}^{+\infty}\frac{d\omega}{2\pi}  
\frac{e^{i {\bf k}\cdot{\bf x}+i\omega t}}{\omega+i \frac{\Gamma}{2} \left(k^2+r\right)}\nonumber \\
&= & -i\Theta(t)  e^{-\frac{\Gamma r }{2}t} \int_{1-\delta \ell <|{\bf k}|/\Lambda<1} \frac{d^dk}{(2\pi)^d} \;  e^{i {\bf k}\cdot{\bf x}-\frac{\Gamma}{2} k^2 t}
\end{eqnarray}
 Taking the limits of $|{\bf x}|\to 0$  and $t\to 0$ we find
\begin{eqnarray}
R^>(0, 0)&=&-i\Theta(0^+)  \int_{1-\delta \ell <|{\bf k}|/\Lambda<1} \frac{d^dk}{(2\pi)^d} =-i\Theta(0^+) \frac{\Omega_d}{(2\pi)^d} \int_{(1-d\ell)\Lambda}^\Lambda\;  dk\;  k^{d-1}  \nonumber \\
&=&-i\alpha \frac{\Omega_d}{(2\pi)^d} \Lambda^d  d\ell  +O(d\ell^2)
\end{eqnarray}
In the last line of this equation,  we have fixed  $\Theta(0^+)=\alpha$,  where $\alpha$ is the stochastic prescription~\cite{Arenas2012-2}. 
Note that the role played by cut-offs is essential to give sense to the limit  $\{|{\bf x}|,t\}\to 0$.  

Differently from tadpole diagrams,   the contributions from the two-loop diagrams of Figure~\ref{fig:2V2loop} are non-local expressions in configuration space.  For instance,  the first diagram in Figure~\ref{fig:2V2loop} is proportional to a functional of $\varphi^<({\bf x}, t)$, given by
\begin{equation} 
I[\varphi^<] = \int d ^dxd^dx'dtdt'\; \varphi^<(\textbf{x},t)[G^>(\textbf{x}-\textbf{x}',t-t')]^3 \varphi^<(\textbf{x}',t')\; 
\label{app:IG3}
\end{equation}
 where
\begin{equation} 
	G^>(\textbf{x}-\textbf{x}',t-t') = \int_{1-\delta \ell <|{\bf k}|/\Lambda<1} \frac{d^dk}{(2\pi)^d}\int_{-\infty}^{+\infty}\frac{d\omega}{2\pi}\;\frac{e^{-i\omega(t-t')+i\textbf{k}\cdot(\textbf{x}-\textbf{x}')}}{\omega^2+\frac{\Gamma ^2}{4}(k^2+r)^2}.\label{app:propagator}
\end{equation}
The external legs represent the fields $\varphi^<$,  located at each vertex ${\bf x}$ and ${\bf x}
'$.   On the other hand, the internal legs  represent the kernel 
$[G^>(\textbf{x}-\textbf{x}',t-t')]^3$, which  connects both vertices.  

This type of momentum-shell two-loop  integrals were used to build the static Renormalization Group in Ref.~\cite{Fradkin-QFTBook}  (please, see Chapter 15, Section 15.2.4 ).   In this Appendix, we describe the method and generalize it to the dynamical case.   The generalization does not introduce any additional difficulty since the time (or frequency) integrals are not constrained. 

It is possible to localize  the  functional $I[\varphi^<]$ (Eq.~\ref{app:IG3}) by expanding  $\varphi^<(\textbf{x}',t')$  around the coordinate $\lbrace\textbf{x},t\rbrace$ as follows:
\begin{equation} 
	\varphi^<(\textbf{x}',t') =\varphi^<(\textbf{x},t) +l_i\partial_i \varphi^<(\textbf{x},t)+\tau\partial_t\varphi^<(\textbf{x},t) +\frac{1}{2} l_i l_j\partial_i\partial_j\varphi^< +... .
\label{app:Taylor}
\end{equation}
where ${\bf l}={\bf x}-{\bf x}'$ and $\tau=t-t'$.  Replacing Eq. (\ref{app:Taylor}) into Eq. (\ref{app:IG3}) we find 
 \begin{eqnarray}    
I[\varphi^<]&=& \Sigma_{3000}\!\!\int d^dxdt\;\varphi^{<2}({\bf x}, t)+\Sigma_{3001}\!\!\int d^dxdt\;\varphi^<({\bf x}, t)\partial_t\varphi^<({\bf x}, t)\nonumber \\ 
&+&\Sigma_{3020}\!\! \int d^dxdt\;\frac{1}{2d}\varphi^<({\bf x}, t)\nabla^2\varphi^<({\bf x}, t)+\dots
\label{app:I3Glocal}
\end{eqnarray}
where the constants
\begin{eqnarray}
 \Sigma_{3000}&=&\int d^dl d\tau\; [G^>({\bf l},\tau)]^3 
 \label{app:Sigma3000} \\
 \Sigma_{3001}&=& \int d^dl\; d\tau \tau [G^>(l,\tau)]^3
 \label{app:Sigma3001} \\
 \Sigma_{3020}&=& \int d^dl d\tau\;  |{\bf l}|^2[G^>(l,\tau)]^3
 \label{app:Sigma3020} 
\end{eqnarray}
We see from Eq.  (\ref{app:I3Glocal}) that $I[\varphi^<]$ is a local functional of $\varphi^<({\bf x}, t)$.  
The ellipsis represents higher order derivatives terms that are irrelevant perturbations at the critical point. 
The notation used for the constants in this expansion  will become clear below. 

To complete this example,   let us show how to compute,   for instance,  $\Sigma_{3000}$ given by Eq. (\ref{app:Sigma3000}).  Using Eq. (\ref{app:propagator}),   Eq.  (\ref{app:Sigma3000}) can be explicitly written as 
\begin{equation}    
	\Sigma_{3000} =\Gamma^3\int_{\rm shell}\frac{d^dk_1d^dk_2d^dk_3}{(2\pi)^{3d}}\int\frac{d\omega_1d\omega_2d\omega_3}{(2\pi)^{3}}\;\frac{ \delta^d\left({\bf k}_1+{\bf k}_2+{\bf k}_3\right)  \delta\left(\omega_1+\omega_2+\omega_3\right)}{\lbrack\omega_1^2+a_1^2\rbrack\lbrack\omega_2^2+a_2^2\rbrack\lbrack\omega_3^2+a_3^2\rbrack} 
\end{equation}	
where $ a_i = (\Gamma/2)(k_i^2+r) $ with $i=1,2,3$.   The integration over ${\bf k}_i$ should be done on the spherical shell    $(1-d\ell)<|k_i|/\Lambda<1$, for $i=1,2,3$.  This fact is indicated by the subscript ``shell" in the momentum integral. 

The frequency integrations can be easily done, using residues theorem,  since they are not constrained.  We get
\begin{equation}
\Sigma_{3000}=\frac{ \Gamma ^3}{4}\int_{\rm shell}\frac{d^dk_1d^dk_2d^dk_3}{(2\pi)^{3d}}\frac{\delta^d\left({\bf k}_1+{\bf k}_2+{\bf k}_3\right)}{ a_1 a_2 a_3 (a_1+a_2+a_3)} \,.
\end{equation}

This integral is extremely difficult (if not impossible) to do exactly.  However, we are interested only in the contribution to order $d\ell$, since this is the only term that enters the recursive DRG equations. 
To do this, we take advantage of the strong constraints imposed by the thin spherical shell ($d\ell\ll 1$) and momentum conservation. The integration domain is composed by the set of all three vectors living in the thin shell  which sum zero. It will be denoted by $D$ and can be sketched by  
\begin{equation} 
\int_D  d^dk_1d^dk_2d^dk_3 = \int_{\rm shell}   d^dk_1d^dk_2d^dk_3 \; \delta^d({\bf k_1}+{\bf k_2}+{\bf k_3})
\label{app:measure}
\end{equation}

We know that the integral is zero in the limit $d\ell \to 0$.  The main goal is to identify,  within the domain $D$,  a sub-domain,  {\em i.e.},  a  non-null set,  that  contributes to the integral at order $d\ell$.   Any other configuration is not interesting for us, since it will not enter the DRG equations.   Thus,  we rewrite the integral in the following way
\begin{equation}
\int_D=\int_{A_{d\ell}}+\int_{A^c} 
\label{app:domainseparation}
\end{equation}
in such a way that $D$ is the domain described before,  $A_{d\ell}$ is the sub-domain of $D$ ($A_{d\ell}\subset D$) which contains all the contributions to order  $d\ell$ and $A^c$ is the complement set, in the sense that $D=A_{d\ell} \cup  A^c$  and  $A_{d\ell} \cap  A^c=\emptyset$.  The leading order of the integral over $A^c$ is, at least,   $d\ell^2$. 

\begin{figure}
	\begin{center}
		\includegraphics[width=0.35\textwidth]{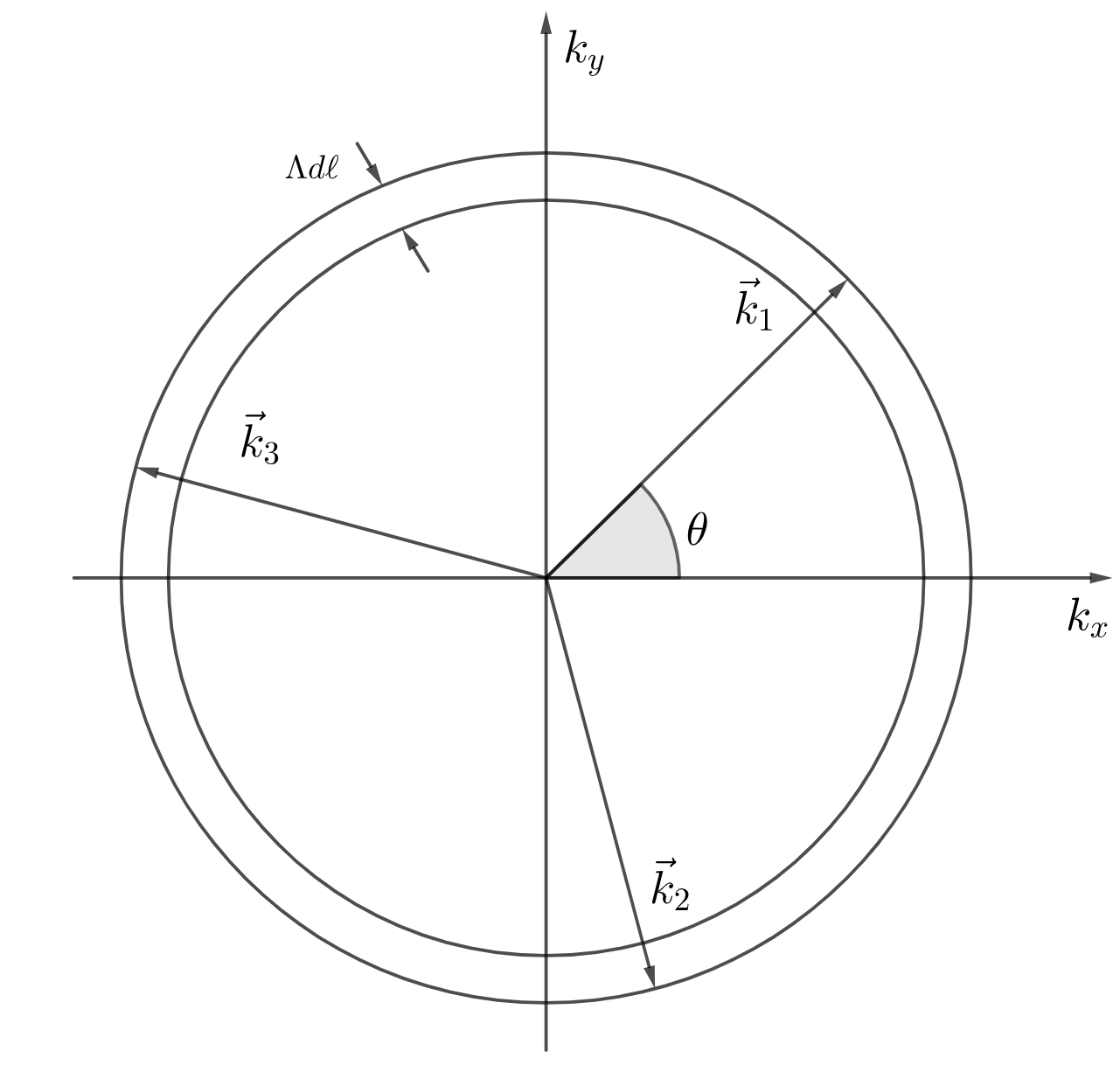}
	\end{center}
	\caption{Typical configuration of internal momenta in a two-loop diagram.  The figure shows the intersection of the plane formed by the three vectors and the d-dimensional spherical shell.   Thus,  the three momenta lie in a thin circular shell $(1-d\ell)<k_i/\Lambda<1$, with $i=1, 2, 3$.  On the other hand,  ${\bf k}_1+{\bf k}_2+{\bf k}_3=0$ due to momentum conservation.}
	\label{fig:constrain}
\end{figure}

To look for  $A_{d\ell}$,  it is convenient to visualize the domain $D$  in a geometrical way.   From a geometrical point of view,  three non co-linear vectors with zero sum are  in the same plane.  Thus,  for a fixed orientation of the plane,  the three vectors live in the intersection of a plane with a d-dimensional spherical shell.  The projection of this intersection on the plain is a circular shell that we depict in Figure~\ref{fig:constrain}.
%%%%%%%%%%%%%%%%%%%%%%%%%%%%%
%%%%%%%%%%%%%%%%%%%
    In the limit of very small $d\ell$,   the vectors form an almost equilateral triangle.  In the limit of $d\ell\to 0$,  the only possible configuration is an exact equilateral triangle inscribed in a circle of radius $\Lambda$.

Let us consider now the set of all equilateral triangles inscribed in circles of radius between   $\Lambda(1-d\ell)<k<\Lambda$,   with $d\ell\ll 1$.  We claim that this set of triangles is our sub-domain $A_{d\ell}$.  In the following, we demonstrate it and compute the integral.  

By definition,  in each configuration of the sub-domain  $A_{d\ell}$ all three momenta have the same modulus 
$k_1=k_2=k_3=k$,   with  $\Lambda(1-d\ell)<k<\Lambda$.   Thus, we can parametrize this set by considering 
\begin{eqnarray}
\int_{\rm shell}  && d^dk_1d^dk_2d^dk_3  \delta^d({\bf k_1}+{\bf k_2}+{\bf k_3}) \times \\
&\times & 
\left\{ \Lambda^2\delta(k_1-k_2)\delta(k_1-k_3)+\left(1-\Lambda^2\delta(k_1-k_2)\delta(k_1-k_3)\right)\right\}
\end{eqnarray}
where we have added and subtracted the term   $ \Lambda^2\delta(k_1-k_2)\delta(k_1-k_3)$.  The factor $\Lambda^2$ was added for dimensional reasons.

We will show that this parametrization is the one we are looking for,   to implement the correct sub-domain $A_{d\ell}$ in such a way that the integration measure corresponding to each part of the integral is
\begin{eqnarray}
\int_{A_{d\ell}} &\to & \Lambda^2   \int_{\rm shell}  d^dk_1d^dk_2d^dk_3  \delta^d({\bf k_1}+{\bf k_2}+{\bf k_3})  \delta(k_1-k_2)\delta(k_1-k_3) 
\label{app:1int}\\
\int_{A^c} &\to   &   \int_{\rm shell}  d^dk_1d^dk_2d^dk_3  \delta^d({\bf k_1}+{\bf k_2}+{\bf k_3}) 
\left\{1- \Lambda^2\delta(k_1-k_2)\delta(k_1-k_3)\right\} 
\label{app:2int}
\end{eqnarray}
It is clear that both integrals have disjoint domains.  The first one is formed by all the momenta with equal modulus that live in the  spherical shell,   while the second one is the complement. 

The integral of Eq.  (\ref{app:1int}) is quite simple to compute.   Firstly, we rewrite the measure in spherical coordinates.  Then,  we simply integrate  over the modulus of $k_1$ and $k_2$ using the modulus $\delta-$functions.  Thus, we are lead with a remaining integral over the modulus of $k_3=k$ and all the angular integrations.   After the integration over $k_1$ and $k_2$,  all the modulus are equal  $k_1=k_2=k_3$.  Therefore,  the d-dimensional $\delta$-function  does not constrain the modulus of $k_3$,  but only constrains the angles in order to guarantee that the vector sum should be zero.   For this reason, there is a remaining integration over the modulus to be done.    On the other hand, the angular integrations are easily visualized geometrically.   The d-dimensional $\delta-$function of the momentum conservation  fixes the angles between the three vectors in exactly $2\pi/3$.  Thus,  the only independent angular degree of freedom  is the orientation of one of the vectors in the plane and the orientation of the plane itself.  Therefore,   the angular integration is that of one single vector over the solid angle of the complete hyper-sphere,  $\Omega_d$.       
To be concrete, 
\begin{eqnarray}
 \Lambda^2\int_{\rm shell}   d^dk_1d^dk_2d^dk_3  \delta^d({\bf k_1}+{\bf k_2}+{\bf k_3}) 
\delta(k_1-k_2)\delta(k_1-k_3) \sim  \Lambda^2  \int d\Omega_d \int_{\Lambda(1-d\ell)}^{\Lambda} dk \;   k^{2d-3}
\label{app:measure-dl}
\end{eqnarray}
Clearly,   this integral will produce a term of order $d\ell$ plus higher order terms. 

The only  point that remains to be elucidated  is the leading order of the integral   over the complement domain $A^c$,  Eq.  (\ref{app:2int}).  In this case, the integration is  performed   over all configurations where the modulus of the three vectors are not equal.   Therefore,   there is no way to reduce this integral to  a single momentum integral.  Fixing the modulus and the orientation of one of the vectors,  say $k_3$,  we need to integrate in the other two momenta $k_1$ and $k_2$.  This contributes to the integral to order at least  $d\ell^2$.  In this way,  the integral over $A_{d\ell}$ has all the contributions to order  $d\ell$ (and of course higher order terms), while the integral over $A^c$ contributes with at least $d\ell^2$ and higher order terms.   
Therefore,   in the limit of an extremely thin shell,  only the first integral  is relevant.    

The  leading order  of the integral over $A_{d\ell}$ is very simple to compute.   In practice,    it is simply to replace all modulus of the momenta  by $\Lambda$ in the integrand  and  finally to perform the angular integrations.   We find 
\begin{equation}
	\Sigma_{3000} = \frac{4}{3\Gamma}\frac{\Lambda^{2d-8}\Omega_d}{(1+r\Lambda^{-2})^4}d\ell + O(d\ell^2)
\label{add:3000}
\end{equation}
where $\Omega_d$ is the area of a $(d-1)$-dimensional sphere of radius one.   This type of two-loop integrals can be found in Ref.~\cite{Fradkin-QFTBook}. 

All the diagrams shown in Figures  from \ref{fig:2V2loop} to \ref{fig:g2_1loop} can be computed,  after the localization process,  as combinations of the integrals
\begin{equation} 
\Sigma_{ijkm}  =  \int d^dl \int d\tau \;G(l,\tau)^i R(l.\tau)^j l^k \tau^m
\end{equation}
where the indices $i, j, k, m$ can take the values $0,1,2,3$.  Equation~\ref{add:3000} is a typical example with $i=3$ and $j=k=m=0$.  We have computed all relevant integrals,  using the methods described above.  The results are: 
\begin{eqnarray}
&&\Sigma_{2000} = \frac{2}{\Gamma}\frac{\Lambda^{d-6}\Omega_d}{(1+r)^3}d\ell 
\hspace{4.4 cm}
\Sigma_{0200} = \frac{-1}{\Gamma}\frac{\Lambda^{d-2}\Omega_d}{(1+r)}d\ell \nonumber
\\
&& \Sigma_{1100} = \frac{-i}{\Gamma}\frac{\Lambda^{d-4}\Omega_d}{(1+r)^2}d\ell 
\hspace{4.2 cm}
\Sigma_{3000} = \frac{4}{3\Gamma}\frac{\Lambda^{2d-8}\Omega_d}{(1+r)^4}d\ell \nonumber
\\
&&\Sigma_{2100} = \frac{-2i}{3\Gamma}\frac{\Lambda^{2d-6}\Omega_d}{(1+r)^3}d\ell 
\hspace{ 3.9 cm}
\Sigma_{1200} = \frac{-2}{3\Gamma}\frac{\Lambda^{2d-4}\Omega_d}{(1+r)^2}d\ell 
\label{app:coef}
\\
&& \Sigma_{2120} = \frac{4i}{9\Gamma}\frac{\Lambda^{2d-8}\Omega_d}{(1+r)^4}\bigg\lbrack d - \frac{4}{3}\frac{1}{(1+r)}\bigg\rbrack d\ell 
\hspace{1.7 cm}
\Sigma_{2101} = \frac{4}{9\Gamma^2}\frac{\Lambda^{2d-8}\Omega_d}{(1+r)^4}d\ell \nonumber
\\
&&\Sigma_{1220} =  \frac{4}{9\Gamma}\frac{\Lambda^{2d-6}\Omega_d}{(1+r)^3}\bigg\lbrack d - \frac{4}{3}\frac{1}{(1+r)}\bigg\rbrack d\ell 
\hspace{1.7 cm}
\Sigma_{1201} = \frac{-4i}{9\Gamma^2}\frac{\Lambda^{2d-6}\Omega_d}{(1+r)^3}d\ell \nonumber
\end{eqnarray}
where we have redefined the constant $r\Lambda^{-2}\to r$, to make it dimensionless. 

%%%%%%%%%%%%%%%
\section{Rescaling of the action}
\label{App:Rescaling}
%%%%%%%%%%%%%%%%%%%
In Section~(\ref{subsec:DRGT}), we described how to obtain the DRG equations (Eqs.  (\ref{eq:r})-(\ref{eq:gamma})).  The main steps were,  after the calculation of all the integrals,  the rescaling of momenta, frequency and fields in $S'_{\Lambda/b}$,  using Eqs. (\ref{eq:bk} -\ref{eq:xiscaled}),  in order to reestablish the original cut-off $\Lambda$ and compare the result with the initial action $S_{\Lambda}$.

In this Appendix we explicitly show how to perform these rescalings.   After the computation of all diagrams we find for the effective action 
\begin{eqnarray}
	S'_{\Lambda/b} &=& \int_{b/\Lambda} dt\; d^d x \bigg\lbrace i\varphi Z_\omega\partial_t \phi -\frac{i\Gamma}{2}\varphi Z_k\nabla^2\phi+\frac{i\Gamma}{2} r Z_r\varphi\phi +\frac{\Gamma}{2}Z_\Gamma\varphi^2 \nonumber
	\\  
	&+&\frac{iu\Gamma}{3! 2}Z_u\varphi \phi^3-\frac{u\Gamma}{4} Z_u\phi^2\bar\xi\xi +\frac{g\Gamma}{2} Z_g\varphi^2\phi^2+ig\Gamma Z_g\varphi\phi\bar\xi\xi\bigg\rbrace  \;
\label{app:correctedAction} .
\end{eqnarray}
where
\begin{eqnarray}
	Z_\omega &= & 1-\frac{u^2\Gamma^2 }{8}\Sigma_{2101}+\frac{iug\Gamma^2}{2}\Sigma_{1201} 
	\\
	Z_k &=&  1+ \frac{u^2\Gamma }{8id}\Sigma_{2120}-\frac{iug\Gamma}{2id}\Sigma_{1220}
	\\
	Z_\Gamma &=& 1 +gG^>(0) +\frac{u^2\Gamma}{3!4}\Sigma_{3000}- iug\Gamma \Sigma_{2100}
	\\
	Z_r &=&  1+\frac{u}{2r}G^>(0) -\frac{2ig}{r}R^>(0) +\frac{u^2\Gamma}{4i r}\Sigma_{2100}-\frac{ug}{r}\Gamma \Sigma_{1200} 
	\\
	Z_u &= & 1-\frac{3}{2}iu\Gamma\Sigma_{1100} -3g\Gamma\Sigma_{0200}
	\\
	Z_g &=&  1- g \Gamma \Sigma_{0200} -\frac{5i u\Gamma}{2}\Sigma_{1100} +\frac{u^2\Gamma}{8g} \Sigma_{2000}  
\end{eqnarray}
with $ \Sigma_{ijkl} $ given by Eq. (\ref{app:coef}). 
Since we know that each $ \Sigma_{ijkl} $ correction (as well as $ G^>(0),R^>(0) $) is of order $ d\ell $, we can simplify the notation and write
\begin{eqnarray}
	Z_\omega &=& 1+ A(r,u,g)d\ell
	\hspace{1. cm}
	Z_k = 1+ B(r,u,g)d\ell
	\\
	Z_\Gamma &=&1+ C(r,u,g)d\ell
	\hspace{1. cm}
	Z_r = 1+ D(r,u,g)d\ell
	\\
	Z_u &=&  1+ E(r,u,g)d\ell
	\hspace{1. cm}
	Z_g =  1+ F(r,u,g)d\ell.
\end{eqnarray}

To find the DRG equations, we have to analyze each term of the actions $S'_{\Lambda/b}$  given by Eq.  (\ref{app:correctedAction}) and compare to $S_\Lambda$.  In the following, we discuss each term separately. 
\begin{itemize}
	\item Regarding the temporal derivative term in \ref{app:correctedAction}, we have in Fourier space, 
	\begin{eqnarray}
		\int_{\Lambda/b}\; \varphi Z_\omega \omega \phi\;   d\omega d^dk	& = \int_\Lambda b^{-d -z +d +2z + \frac{\eta_2-\eta_1}{2}-z}\; Z_\omega\varphi\omega\phi\;  d\omega d^dk \; .
	\end{eqnarray}
	Comparing with $S_\Lambda$ we get
	\begin{eqnarray} 
			Z_\omega\; b^{ \frac{\eta_2 -\eta_1}{2}} = 1. \nonumber
	\end{eqnarray}
	Since $b = 1+d\ell$,    the difference between anomalous dimensions $\eta_2-\eta_1$ is given by
	\begin{eqnarray}
		\frac{\eta_2 -\eta_1}{2} = \frac{Z_\omega^{-1} -1}{d\ell} = -  A(r,u,g)
		 \label{app:eta1eta2}
	\end{eqnarray}
	
	\item The second term of Eq.  (\ref{app:correctedAction}) proportional to the Laplacian of $\phi$ scales (in Fourier space) as 
	\begin{eqnarray}	
	\int_{\Lambda/b}\;  \frac{\Gamma}{2}\varphi Z_k k^2\phi\; d\omega d^dk	 =	\int_\Lambda b^{-d -z +d +2z + \frac{\eta_2-\eta_1}{2}-2}\;\frac{\Gamma}{2} Z_k\varphi k^2\phi\; d\omega d^dk
\end{eqnarray}		
Comparing with $S_\Lambda$ we get		
		\begin{eqnarray}
		\Gamma'= \Gamma   Z_k\; b^{z-2 +\frac{\eta_2 -\eta_1}{2}}  =  \Gamma   Z_k   Z_\omega^{-1}\;b^{z-2 }
\label{app:Gammaprime}	
	\end{eqnarray}
	Considering the infinitesimal transformation $b=1+d\ell$ we find 
	\begin{equation}
		\frac{d\Gamma }{d\ell}= \Gamma   (z - 2 +B +\frac{\eta_2 -\eta_1}{2} )	
	\end{equation}
	or, using Eq.  \ref{app:eta1eta2},
	\begin{equation}
		\frac{d\Gamma }{d\ell}= \Gamma   (z - 2 +B(r, u, g) -A(r, u, g) )
	\end{equation}
Replacing in this expression the explicit forms of $A(r, u, g)$ and $B(r, u, g)$
  we find Eq.   (\ref{app:gamma}).
  
	\item Eq \ref{app:r} can be found from the scaling of the third term of Eq. (\ref{app:correctedAction}):
	\begin{eqnarray}
		\int_{\Lambda/b}\;\frac{\Gamma r}{2} Z_r\varphi\phi	\; d\omega d^dk	 =	\int_\Lambda b^{-d -z +d +2z + \frac{\eta_2-\eta_1}{2}}\;\frac{\Gamma r}{2} Z_r\varphi \phi\; d\omega d^dk	
	\end{eqnarray}
Comparing with $S_\Lambda$, 
\begin{eqnarray}	
		\Gamma' r'= \Gamma r  Z_r\; b^{z +\frac{\eta_2 -\eta_1}{2}} 
\end{eqnarray}
Using the explicit form of $\Gamma'$ given by Eq. (\ref{app:Gammaprime})
\begin{eqnarray}
  r'=  Z_k^{-1}   Z_\omega  Z_r\;r b^{2+\frac{\eta_2 -\eta_1}{2}}  =  Z_k^{-1}   Z_r\;r b^{2}
	\end{eqnarray}
and considering the infinitesimal transformation $b=1+d\ell$ we finally obtain
	\begin{equation}
		\frac{dr }{d\ell}= r  (2 -B(r, u, g) +D(r, u, g) )\; .
	\end{equation}
By replacing the explicit expressions for $B(r, u, g)$  and $D(r, u, g)$  we find Eq. (\ref{app:r}).
	
	\item  Regarding the $u$ coupling, we have:
	\begin{eqnarray}
		\int_{\Lambda/b}\; d\omega d^dk\;\frac{u\Gamma}{3! 2}Z_u\varphi \phi^3	 =	\int_\Lambda\; d\omega d^dk\; b^{-3d -3z +2d +4z +2+ \frac{3\eta_2}{2}-\frac{\eta_1}{2}}\;\frac{u\Gamma}{3! 2}Z_u\varphi \phi^3
	\end{eqnarray}
After comparing with $S_\Lambda$  and following the same procedures we find
\begin{eqnarray}		
		 u'= u  Z_k^{-1} Z_u\; b^{ 4-d +\eta_2}
	\end{eqnarray}
Considering infinitesimal transformations  and the expressions for $Z_k$ and $Z_u$ we get
		\begin{equation}
		\frac{du }{d\ell}= u  (4-d +\eta_2 + E(r, u, g)-B(r, u, g) );
	\end{equation}
that is equivalent to Eq \ref{app:u} after replacing the explicit values of $E(r, u, g)$ and $B(r, u, g)$.

	\item Finally, for Eq. \ref{app:g} we need to rescale 
	\begin{eqnarray}
		\int_{\Lambda/b}\;\frac{g\Gamma}{2} Z_g\varphi^2\phi^2\; d\omega d^dk\;	 =	\int_\Lambda b^{-3d -3z +2d +4z + \eta_2-\eta_1}\;\frac{g\Gamma}{2} Z_g\varphi^2\phi^2\; d\omega d^dk\;
		\end{eqnarray}  
		in such a way that 
		\begin{eqnarray}
		\Gamma'g'= \Gamma g  Z_g\; b^{z -d +\eta_2-\eta_1} 
		\end{eqnarray}
Using  previous results we easily find 		
		\begin{eqnarray}
			  g'= g  Z_k^{-1}Z_\omega Z_g\; b^{2-d+\eta_2-\eta_1} = g  Z_k^{-1}Z_\omega^{-1} Z_g\; b^{2-d}
	\end{eqnarray}
	Therefore, we arrive at the DRG equation for the $g$ coupling as
	\begin{equation}
		\frac{dg }{d\ell}= g  (2-d +\eta_2-\eta_1 + A(r, u, g)-B(r, u, g)+F(r, u, g) )
	\end{equation}
	or by using Eq.  (\ref{app:eta1eta2})
	\begin{equation}
		\frac{dg }{d\ell}= g  (2-d -A(r, u, g)-B(r, u, g)+F(r, u, g) )
	\end{equation}
	which coincides with Eq. (\ref{app:g}) after replacing explicit expressions for $A(r, u, g)$,  $B(r, u, g)$ and $F(r, u, g)$.
\end{itemize}

\end{document}